\documentclass[12pt,a4paper]{article}
\usepackage{amsmath,amssymb,amsthm,amscd, graphicx, color, verbatim, slashed}

\usepackage{mathrsfs}

\usepackage{graphicx}

\usepackage[all]{xy}

\usepackage{tikz}
\usetikzlibrary{matrix,arrows}

\makeatletter

\@addtoreset{equation}{section}
\makeatother

\newcommand{\dpa}[1]{\frac{\partial}{\partial #1} }

\newcommand{\mbb}{\mathbb}
\newcommand{\Oo}{\mathcal{O}}
\newcommand{\xto}{\xrightarrow}
\newcommand{\mf}{\mathfrak}
\newcommand{\ip}[1]{\left\langle #1 \right\rangle}
\newcommand{\beq}{\begin{equation}}
\newcommand{\eeq}{\end{equation}}

\newcommand{\cinfty}{C^{\infty}}
\newcommand{\eps}{\epsilon}
\renewcommand{\d}{\mathrm{d}}
\newtheorem*{prop*}{Proposition}

\newtheorem*{note*}{Note}

\newtheorem*{ques*}{Question}
\theoremstyle{remark}

\theoremstyle{definition}

\newcommand{\Z}{\mathbb{Z}}
\newcommand{\R}{\mathbb{R}}
\newcommand{\C}{\mathbb{C}}

\newcommand{\A}{\mathcal{A}}

\newcommand{\del}{\partial}

\newcommand{\abs}[1]{\left\lvert #1 \right\rvert}

\newcommand{\zbar}{{\bar{z}}}

\newcommand{\what}{\widehat}

\newcommand{\nc}{\newcommand}
\nc{\mc}{\mathcal}
\newcommand{\scB}{\ensuremath{\mathcal{B}}}

\theoremstyle{definition}

\newcommand{\dbar}{\overline{\partial}}
\newcommand{\br}[1]{\overline{#1}}
\newcommand{\Q}{\mathbb{Q}}
\newcommand{\op}{\operatorname}

\begin{document}

\begin{titlepage}
\hfill \\
\vspace*{15mm}
\begin{center}
{\Large \bf Topological Chern-Simons/Matter Theories}

\vspace*{15mm}

{\large Mina Aganagic$^a$, Kevin Costello$^b$, Jacob McNamara$^c$ and Cumrun Vafa$^{c}$}
\vspace*{8mm}

${}^a$ Center for Theoretical Physics, University of California, Berkeley\\
\vspace*{0.2cm}
${}^b$ Perimeter Institute for Theoretical Physics\\
\vspace*{0.2cm}
${}^c$ 
Jefferson Physical Laboratory, 
Harvard University\\

\vspace*{0.7cm}

\end{center}
\begin{abstract}

We propose a new partially topological theory in three dimensions which couples Chern-Simons theory to matter.  The 3-manifolds needed for this construction admit transverse holomorphic foliation (THF).  The theory depends only on the choice of such a structure, but not on a choice of metric and in this sense, it is topological.
We argue that this theory arises in topological A-model string theory on Lagrangian 3-branes in the presence of additional parallel coisotropic 5-branes.  The theory obtained in this way is equivalent to an ${\cal N}=2$ supersymmetric Chern-Simons matter theory on the same 3-manifold, which also only depends on the THF structure.  The theory is a realization of a topological theory of class H, which allows splitting of a temporal direction from spatial directions.  We briefly discuss potential condensed matter applications.

\end{abstract}

\end{titlepage}




\section{Introduction}
Topological Chern-Simons theory in three dimensions \cite{Witten:1988hf} is an example of a rich QFT which nevertheless depends only on the topology of the 3-manifold on which it is defined.  It has applications to 3-manifold and link invariants, but also to condensed matter systems.  Quite surprisingly it was shown in \cite{Witten:1992fb} that this theory can also be viewed as the target space theory on topological A-model D-branes in Calabi-Yau manifolds of the form $X=T^*M^3$, where D-branes are wrapped on the Lagrangian $M^3$ in $X$.

However, it was noticed in \cite{Kapustin:2001ij} that one can also in principle introduce 5-dimensional coisotropic branes in such a setup.  Namely, they showed that the topological A-model admits 5-branes, in addition to 3-branes. In a compact Calabi-Yau with $SU(3)$ holonomy, 5-cycles to support such branes do not exist, but in a non-compact Calabi-Yau they do, as we will see. As shown in \cite{Kapustin:2001ij, Kapustin:2003kt} the 5-dimensional submanifold should admit a transverse holomorphic structure, which means that there is a given 1-dimensional foliation (viewed as `time') as well as an integrable complex structure on the tangent bundle modulo the foliation.  It is natural to ask what theory lives on this 5d defect and moreover what theory describes the combined 3d/5d system. We will show that, when the 3-brane is a subspace of the 5-brane, and the flux on the 5-brane vanishes when restricted to the 3-brane, the 3-brane inherits the transverse holomorphic fibration (THF) structure. The strings stretched between the 3-brane and 5-brane lead to matter charged under the gauge symmetry.  We thus expect to have CS theory coupled to matter.  Moreover, from the setup alone, it is clear that it has to be a partially topological theory, in that it will depend on the THF structure but not on metric on the manifolds.

On the other hand, in the study of 3-dimensional ${\cal N}=2$ supersymmetric field theories, it was shown in \cite{Klare:2012gn, Closset:2012ru} that one can preserve at least one supercharge on 3-manifolds which admit a THF structure.
Moreover the partition function only depends on moduli of the THF structure, and not on the metric on the manifold.
So in this sense these theories also lead to partially topological 3d theories.

We argue that the ${\cal N}=2$ supersymmetric field theories on THF manifolds and the 3d/5d system arising
in topological string are identical physical systems (viewing the 5d coisotropic brane as non-compact source).  
We provide a 3d Chern-Simons theory coupled to matter which describes the ${\cal N}=2$ theories on THF manifolds, and the topological string 3d/5d system. It is a partially topological theory that only depends on THF structure.  We propose a formula for its partition function, by integrating out the matter fields. This leads to certain combination of link invariants in the standard Chern-Simons theory, which is readily computable. The formula involves contributions from closed leaves of the foliation. 

The THF can be viewed more physically as picking out a time evolution direction in the three manifold.  In other words, the space locally looks like $(t,z)$ where $t$ is the real `time' coordinate and $z$ is complex. In a condensed matter setup, where Lorentz invariance is not built in, such an effective system can arise in the IR. This gives a concrete example of a `class H' topological theory \cite{Kong:2014qka, Freed:2016rqq}: The theory is topological, except for this splitting of space and time. It is natural to conjecture that in some condensed matter system this QFT can describe the IR physics of the dynamical anyons coupled to Chern-Simons theory.

The organization of this paper is as follows:  In section 2 we begin by first defining what is meant by a THF structure on a manifold. We then define the action for CS theory coupled to matter which only depends on the THF structure. We also give a heuristic argument for the proposed localization of the partition function of this theory to closed "time-like" orbits. In section 3, we explain how this theory arises in the context of topological strings on CY 3-folds with Lagrangian and coisotropic branes. 
In section 4, we study the topological string field theory in the special case when the Calabi-Yau is a product manifold, and show it exactly reproduces the Chern-Simons matter theory from section 2.  In section 5, we prove that Chern-Simons matter theory we defined arises from 3d ${\cal N}=2$ gauge theories, with topological twisting depending on the TFH structure. In section 6, we consider two examples corresponding to $M^3=S^3$ and $S^2\times S^1$.  In particular for the latter case, we give a rigorous derivation of the proposed partition function for a specific THF on $S^2\times S^1$.  Moreover we show that in these two cases we recover the partition function of ${\cal N}=2$ supersymmetric matter coupled to CS theory on these two spaces.  Finally, in section 7, we point to a possible condensed matter application.

\section{A partially topological CS/matter theory}

In this section we define a 3d topological theory which involves Chern-Simons theory coupled to matter.  The resulting theory is not defined on an arbitrary 3-manifold, but only on ones which admit `transverse holomorphic foliation' (THF). It is partially topological in that it only depends on choosing a THF and not on any other data such as a metric on the manifold.  In order to define our fields and the action, we need to review some preliminaries about THF manifolds first.

\subsection{3-manifolds with THF}
A 3-manifold admits a THF if we can choose local coordinates $(t,z)$ where $t\in \mathbb{R}$ and $z\in \mathbb{C}$.  Moreover, as we go from patch to patch the transformations are of the type:
\begin{equation}\label{coordinates}
t'(t,z,{\overline z}),\quad z'(z), \quad {\overline z'}({\overline z}).
\end{equation}
Compact 3-manifolds with THF have been classified \cite{classified1, classified2, classified3} and admit a finite number of deformation parameters each (see \cite{Closset:2012ru}, section 5, for a more details). They are all either Seifert manifolds, i.e. circle fibrations over a Riemann surface, or $T^2$ fibrations over $S^1$. 
In this sense 3-manifolds with THF are analogous to Riemann surfaces, whose space of complex structures is finite dimensional ($3g-3$ complex dimensional for a Riemann surface of genus $g$).

The form (\ref{coordinates}) for the coordinate changes means in particular that the space of holomorphic 1-forms $\Omega^{1,0}$ on a THF manifold is well-defined.  Similarly, we denote by $\Omega^{0,1}$ the space of anti-holomorphic $1-$forms. Note that the $dt$ direction is not well defined, because of the mixing with $(z,{\overline z})$ as we go from one patch to another.  The quotient spaces $\Omega /\Omega^{1,0}$ and $\Omega/\Omega^{0,1}$ are well-defined, where $\Omega$ is the space of all 1-forms, as is the projection from $p:\Omega \rightarrow \Omega/\Omega^{1,0}$, which simply means:
$$p:\qquad fdt+g dz +h d{\overline z}\rightarrow fdt+hd{\overline z},$$
in local coordinates. Using this projection operator, we can define a modified $d$ operator as 
$$
\hat d=p\cdot d.$$  
In coordinates:
$$\hat df= {\partial f\over \partial t}dt+{\partial f\over \partial {\overline z}}d{\overline z}.$$
We are now ready to define the theory.

\subsection{Partially topological CS/matter action}
The fields in the theory are $\phi \in \Omega^{1,0}\otimes R$ and $\eta \in \Omega/\Omega^{1,0}\otimes R^c$, and $A$
where $R$ denotes a representation of gauge group $G$, $R^c$ is the conjugate one, and $A$ is the corresponding gauge field.  

We define the action by
$$S={ik\over 4\pi}\int CS(A)+\int \eta \wedge d_A \phi.$$ 
Note that there is an additional symmetry, which we view as a gauge symmetry, given by:
\begin{align}\label{gta}
 \eta &\rightarrow \eta + {\hat d_A} \epsilon \\
 A & \rightarrow A + \frac{4 \pi i }{k} \mu (\phi,\epsilon)
 \end{align}
where $\epsilon \in R^c$ is an arbitrary function of spacetime, 
$$\hat d_A= p\cdot (d-i A),
$$ 
and 
$$
\mu : R \otimes R^c \to \mf{g}
$$
is the moment map. 

The interaction of matter, represented by $\eta,\phi$ can be written in local coordinates as
\begin{equation}\label{matter}
\int (\eta_{\overline z}\, \del_t \phi_z -\eta_t\,{\del_{\overline z}}\phi_z)\ dt \wedge dz \wedge d\zbar,
\end{equation}
when $A = 0$, leading to equations of motion:
$$\del_t \phi_z=0,\qquad \del_{\overline z} \phi_z=0,\qquad \del_t \eta_{\overline z}-\del_{\overline z} \eta_t=0.$$
Thus, on-shell, we can view $\phi_z$ as time-independent and holomorphic.
Using the equations of motion for $\eta$, by a gauge choice (choosing a suitable $\epsilon$), we can locally set $\eta=0$.

Before we go on, note that the matter system can be made massive in a way that preserves the partial topological invariance, and gives the ${\phi, \eta}$ system a real mass $m$. We introduce a background complex connection $A_b$, by replacing 
$$A\; \rightarrow \; A - A_b,$$ 
in the matter action. As we will see later, $A_b$ has string theory interpretation as the connection on the 5-brane. The $dt$ component of it can be made complex, since the imaginary part of it is the one real modulus for the 5-brane position in the Calabi-Yau.

\subsection{The theory at the level zero}
There is an interesting variant of this construction that arises when we take a certain scaling limit of Chern-Simons theory at level zero.\footnote{More properly, because of the one-loop shift in the level of Chern-Simons theory, this should be called the critical level.  For now we will stick with the terminology derived from considering the classical Chern-Simons Lagrangian.}  Naively, the limit of the Chern-Simons action
\begin{equation}
\frac{i k }{4 \pi}\int CS(A) 
\end{equation}
as $k \to 0$ does not lead to a well-defined theory. However, if we are on a 3-manifold with a transverse holomorphic fibration, we can perform a scaling of the fields as $k \to 0$ which will give us a well-defined limit.

Let us decompose our gauge field as 
\begin{equation}
A = A^{1,0} + A'
\end{equation}
where $A^{1,0}$ is, locally, the $\d z$ component and $A'$ is locally the $\d \zbar$ and $\d t$ components.  The Chern-Simons action then looks like
\begin{equation}
\frac{i k }{4 \pi}\int \left( A^{1,0} F(A') + \tfrac{1}{2} A' \d A'  \right) 
\end{equation}
If we perform the change of coordinates
\begin{equation}
B = \frac{ik}{4 \pi} A^{1,0}
\end{equation}
then the Chern-Simons action becomes 
\begin{equation}
\int  B F(A') + \frac{i k }{8 \pi}\int A' \d A' . 
\end{equation}
This clearly has a well-defined $k \to 0$ limit, where we drop the second term.

In this limit, we view
\begin{align}
B &\in \Omega^{1,0} \otimes \mf{g} \\
A' & \in \Omega^1/\Omega^{1,0} \otimes \mf{g}.
\end{align}
Then  $F(A')$ is the curvature of $A'$ modulo any terms involving $\d z$:
\begin{equation}
F(A)' = \what{d} A' + \tfrac{1}{2} [A', A'] \in \Omega^2 / (\Omega^{1,0} \wedge \Omega^1) \otimes \mf{g} . 
\end{equation}
Taking the $k \to 0$ limit also changes the gauge symmetry, because it involves rescaling   the $\d z$ component of $A$. In the limit, $B$ transforms as an adjoint-valued section of a line bundle, and not as part of a connection.

We can couple $k \to 0$ Chern-Simons theory to matter, leading to a theory with action
\begin{equation}
\int B F(A') + \int \phi \,\d_{A'} \eta
\end{equation}
where, as before, $\phi \in \Omega^{1,0} \otimes R$ and $\eta \in \Omega^1/\Omega^{1,0} \otimes R^c$. 
If $\eps$ is an infinitesimal generator of the gauge symmetry of the matter fields, it acts by
\begin{align}
\eta & \mapsto  \eta + \hat{d}_A \eps\\
B & \mapsto B - \mu ( \phi,\eps)
\end{align}
where $\mu : R \otimes R^c \to \mf{g}$ is the moment map.
 
As we will see later, the Chern-Simons matter theory we are studying can be seen as a partially topological twist of ${\cal N}=2$ supersymmetric gauge theory with matter and with a Chern-Simons term.  The level zero limit discussed in this section is simply the twist of the same ${\cal N}=2$ gauge theory without a Chern-Simons term.

\subsection{BV-BRST }\label{BV-BRST}

To complete the full gauge-fixed version of the action, we work in the BV-BRST formalism. In addition to the fields
\[ \phi \in \Omega^{1, 0} \otimes R, \quad \eta \in (\Omega/\Omega^{1, 0}) \otimes R^c, \]
we had before, we also introduce a ghost
\[ \chi \in R^c, \]
for the gauge transformation of $\eta$, and anti-fields and anti-ghosts
\[ \phi^\vee \in \Omega^2/(\Omega^{1, 0} \wedge \Omega^1) \otimes R^c, \quad \eta^\vee \in (\Omega^{1, 0} \wedge \Omega^1) \otimes R, \quad \chi^\vee \in \Omega^3 \otimes R. \]
In local coordinates, we have
\[ \phi^\vee = \phi^\vee_{\overline{z} t} d\overline{z} \wedge dt, \quad \eta^\vee = \psi_{z t} dz \wedge dt + \psi_{z \overline{z}} dz \wedge d \overline{z}, \quad \chi^\vee = \chi^\vee_{z \overline{z} t} dz \wedge d\overline{z} \wedge dt. \]
We will find it convenient to write
\begin{equation}\label{BV fields}
\begin{split}
\Phi &= \phi + \eta^\vee + \chi^\vee \in (\Omega^{1, 0} \wedge \Omega^*) \otimes R, \\
\Psi &= \chi + \eta + \phi^\vee \in \Omega^*/(\Omega^{1, 0} \wedge \Omega^*) \otimes R^c[1].
\end{split}
\end{equation}
The symbol $[1]$ indicates a shift in ghost number, so that the $k$-form component of $\Psi$ is in ghost number\footnote{The convention we are using here is that a physical field has ghost number zero, an anti-field has ghost number $1$, a ghost for a symmetry has ghost number $-1$, and so on.  When one considers local operators built from the various fields, this grading gets reversed: so that a local operator which is a linear function of a ghost field has ghost number $1$. In this convention the  BRST operator always has ghost number $+1$.} $k-1$. 

We will also introduce the full BV-BRST field content of Chern-Simons theory. The ghosts, fields, anti-fields and anti-ghosts for Chern-Simons arrange into a single inhomogeneous differential form
\begin{equation}
\A \in \Omega^\ast \otimes \mf{g}[1].
\end{equation}
The BV-BRST form of the Chern-Simons Lagrangian is given by
\begin{equation}
CS(\mathcal{A}) = \tfrac{1}{2} \ip{\mathcal{A}, \d \mathcal{A}} + \tfrac{1}{6}  \left\langle\mathcal{A},[\mathcal{A},\mathcal{A}] \right\rangle. 
\end{equation}
The action is then given by
$$S = \frac{ik}{4 \pi} \int CS(\mathcal{A}) + \int \Psi \wedge d_\A \Phi.$$ 
It is a formal consequence of the fact that gauge symmetry holds off-shell that this action satisfies the classical master equation.

Note that if we expand this action out, there is a term of the form $\int \chi \phi A^\vee$, where $\chi$ is the ghost for the gauge symmetry in the matter sector and $A^\vee$ is the anti-field for the Chern-Simons gauge field.  This term reflects the fact that the gauge field $A$ varies as in \eqref{gta}.
under a gauge transformation in the matter sector whose generator is $\eps$.

\subsection{The level zero theory in the BV-BRST formalism}
The level zero theory has a very similar description in the BV-BRST formalism.  We can introduce fields
\begin{align} 
\A' &\in  \Omega^\ast / (\Omega^{1,0} \wedge \Omega^\ast) \otimes \mf{g} [1] \\
\scB & \in \Omega^{1,0} \wedge \Omega^\ast  
\end{align}
which encode the fields, ghosts, anti-fields, and anti-ghosts of level zero Chern-Simons. The action functional for the level zero theory with matter is 
\begin{equation}
\int \scB \d \A' + \tfrac{1}{2} \int \scB [\A',\A'] + \int \Psi \d \Phi + \int \Psi \A' \Psi.
\end{equation} 

\subsection{Integrating out Matter}\label{integrate_out_matter}

Consider the path-integral for the matter action \eqref{matter} for some compact THF structure.  We will view
$t$ as time, and compute the path integral in operator formulation. 
 
To begin with, note that $\eta_{\overline z}(t,z, {\overline z})$ is canonical conjugate variable to $\phi_z (t,z, {\overline z})$. Performing the integral over $\eta_t(t, z, {\overline z})$, inserts a delta function in the path integral, localizing it to configurations where ${\overline D}_{\overline z} \phi_z=0$. Thus, ${\phi_z}$ is the only degree of freedom left, and moreover it depends on $z$ only holomorphically,
$$
 {\phi_z} = \phi_z(t,z).
$$
(The dependance of $\eta_{\overline z}$ on ${\bar z}$ can be gauged away using $\eta_{\overline z} \rightarrow \eta_{\overline z} + \partial_{\overline z} \epsilon$.) 

Note that the local Hamiltonian in this framework vanishes, 
$H=L-pdq/dt=0$, as expected for a topological theory. The only contributions the Hamiltonian come from the (background) connections $A$ and $A_b$.

\subsubsection{Localization to closed orbits}

We will first argue that only the modes along closed leaves of foliation contribute to the partition function in any generic enough situation. 

Assume the background complex connection is not zero, 
$$A_b = i\,  m \, dt+\ldots,$$
which is a generic situation.  The term in the action $\int \eta_{\overline z}\, i A_b \,\phi_z$ contributes to the Hamiltonian $H$ and leads to a constant contribution to the energy of all the modes, proportional to $m$. 

Consider a leaf ${{\cal C}}_{\alpha}$ of the foliation along which 
$$\oint_{{\cal C}_{\alpha}} A_b = \beta_{\alpha}.
$$ 
If ${\cal C}_{\alpha}$ is a closed leaf $\beta_{\alpha}$ is finite, and otherwise not. Modes along any orbit be suppressed by $\exp(-\beta_{\alpha} H)$, which kills the contribution of all but the closed leaves of foliation.

\subsubsection{Contribution of a closed leaf }
We will assume we are in a situation where the closed leaves of foliation are isolated, and determine the contribution to the partition function of such leaves. 

Consider the leaf ${{\cal C}}_{\alpha}$ corresponding to a closed orbit, at $z=0$. We can write a complete set of states for the Hilbert space, near $z=0$,  by considering 
$$\phi_z=\sum_{n\geq 0}\phi_n z^n.
$$  
The Hilbert space ${\cal H}({\cal C}_{\alpha})$ corresponding to the leaf is given by the Fock space, generated by $\phi_n$'s acting on the vacuum.   

 As we go around the circle, the THF requires $z$ comes back to itself, up to some rotation $q_\alpha$. This corresponds to having an additional connection
$$
D_t  \rightarrow D_t+ ( b_{\alpha} z d/dz + cc) 
$$ 
be part of the action, where $q_{\alpha}= \exp(b_{\alpha})$. In the operator formalism, it leads to
insertion of $q_{\alpha}^J$ in the contribution of this orbit to the partition function:

$$
{\rm Tr}_{{\cal H}({\cal C}_{\alpha})}  \bigl( \;q_{\alpha}^J\cdot U_{\alpha} \cdot U_{b, \alpha}^{-1}  \bigr),
$$
where $J= z{d\over dz}$. Above, $U_{\alpha}$ is the holonomy of the Chern-Simons connection $A$ around ${\cal C}_\alpha$, and 
$U_{b, \alpha}$  is the holonomy of the complexified background connection $A_b$, 
$$
U_\alpha = e^{i \oint_{{\cal O}_{\alpha}} A}, \qquad U_{b,\, \alpha} = e^{i \oint_{{\cal O}_{\alpha}} A_b}= e^{\beta_{\alpha} m}.
$$
Using this we find  the contribution of the orbit to the partition function:
\beq\label{oo}
Z_{matter}({\cal C}_\alpha; U_{\alpha}) =  \prod_n (1-q_\alpha^n  \; U_\alpha \;  e^{-\beta_{\alpha} m}).
\eeq
%
To get the full partition function of the theory, we simply view $Z_{matter}$ as an operator insertion in the underlying CS theory of the Lagrangian three brane, and take the product over all the finite leaves. The final partition function can be evaluated by computing
\beq\label{eoo}
\langle  \prod_{\alpha} Z_{matter}({\cal C}_\alpha; U_\alpha)\rangle_{CS}
 \eeq
 The closed leaves are  isolated and the product over $\alpha$ is finite, by the genericity assumption.

\section{ Lagrangian and Coisotropic Branes in Topological String}

In this section, we will explain how to obtain, from topological string theory, the Chern-Simons-matter system on a three manifold $M^3$ with THF structure. The result will be a $U(N)$ Chern-Simons theory on $M^3$ with an $(\eta, \phi)$ multiplet in fundamental representation, or with any number $k$ of copies of the fundamental matter multiplet, when the theory gets an additional $U(k)$ global symmetry.
\subsection{Chern-Simons Theory from Lagrangian Branes}

To get a $U(N)$ Chern-Simons theory on a three manifold $M$ in string theory, we start with the 
topological A-model on a Calabi-Yau $X$, which is the total space of the cotangent bundle to $M$:

$$
X=T^*M.
$$
Topological A-model admits 3-branes supported on any Lagrangian submanifold $L$ of $X$. A Lagrangian sumbanifold of $X$ is a 3-real dimensional manifold (which half the dimension of $X$) on 
which the restriction of the symplectic form $\omega$ vanishes. A Lagrangian 3-brane has additional data, namely a connection
$A$ which is flat,
\beq\label{flat}
 F=0,
\eeq
and valued in the Lie algebra of $U(N)$, where $N$ is the number of branes on $L$. 
The flat connection $F=dA0$ is the critical point of the Chern-Simons action functional. For general $(L,X)$, there are worldsheet instanton corrections, coming from holomorphic maps to $X$ with boundaries on $L$ which make the theory more complicated. If, however, we take $(L,X)=(M, T^*M)$,  then \eqref{flat} is exact.

With $N$ 3-branes on $M$ in $X$, it was shown in \cite{Witten:1992fb} that the theory on $M$ is exactly the $U(N)$ Chern-Simons theory:
\beq\label{CS}
S_{CS} = {i k \over 4 \pi} \int_M CS(A) = {i k \over 4 \pi} \int_M {\rm Tr}(A\wedge  dA + {2\over 3} A\wedge A\wedge A).
\eeq
The Feynman diagrams of Chern-Simons theory coincide with open topological string diagrams with boundaries on $M$ so that \eqref{CS} is the string field theory action of $N$ A-branes on $M$.

\subsubsection{Additional Lagrangian Branes}
We can introduce additional Lagrangian A-branes in this setup. A well known case, see \cite{Aganagic:2013jpa} for a review, corresponds to taking the Lagrangian $L_K$ to be the co-normal bundle to the knot $K$ in $M$. The effect of this on the  Chern-Simons theory on $M$ can be derived by integrating out open strings with one boundary on $M$ and one on $L_K$. This leads to the following observable
$$
{\rm det}(1 - e^{-2\pi m} U_K \cdot U_b^{-1}),
$$ 
$U_K$ is the holonomy of the Chern-Simons gauge field along the knot $K$.   $U_b \cdot e^{2\pi m}$ is the complexified holonomy of the background gauge field on $L_K.$ The parameter $m$ corresponds to the real modulus of the Lagrangian brane on $L_K$, which allow us to push the brane off of the $M$ and give mass $m$ to the strings stretching between $M$ and $L_K$. This gets paired up with the holonomy of the real gauge field $A_b$ on $L_K$ around the knot, in one complex modulus.

Introducing additional branes on a Lagrangian $L\neq M$ in $X$ necessarily breaks the symmetries of the vacuum Chern-Simons theory on $M$. By contrast, we can preserve almost all of the topological symmetry of $M$ by adding to $X$ coisotropic $A$-branes instead.

\subsection{Coisotropic A-branes on $X$}
In addition to Lagrangian A-branes which are 3-branes in $X$, the A-model admits boundaries on 5-branes supported on a coisotropic sub-manifold $Y$ of $X$ \cite{Kapustin:2001ij,Kapustin:2003kt}.  As explained in \cite{Kapustin:2001ij}, the curvature $F$ of the gauge field on the coisotropic brane needs to satisfy certain conditions which give the coisotropic submanifold a THF structure. THF structure in 5 dimensions means that $Y$ admits local coordinates $(z_1,z_2,t)$ where $z_i$ is complex and $t$ is real, and in going from one patch to another the coordinates mix holomorphically in transverse directions, i.e. $(z_1'(z_1,z_2) ,z_2'(z_1,z_2),t'(t,z_i,\overline{z_i}))$. The transverse holomorphic structure should not be confused with the complex structure coming from $X$.

\subsubsection{Definition of Coisotropic Branes}
A coisotropic submanifold $Y$ of dimension 5 is a level set of a some real valued function ${\cal H}$ on $X$:
$$
{\cal H} =m
$$
where $m$ is a real parameter, a modulus of $Y$. We automatically also get a vector field $v$ on $Y$, defined by
\beq\label{Ham}
i_v\omega = d{\cal H}.
\eeq
Here, $v$ can be viewed as $v={\partial/\partial t}$.
If we view $X$ as a phase space with symplectic form ${\omega}$ and ${\cal H}$ as a Hamiltonian, then $v$ is the vector field corresponding to `time' translations generated by ${\cal H}$. 

A coisotropic brane wrapping $Y$ needs to carry a curvature two-form $F$ such that
\beq\label{first}
i_v F=0
\eeq
viewed as one form on $Y$.   Moreover, $F$, on the brane, must satisfy
\beq
F\wedge \omega =0, \qquad F\wedge F = \omega\wedge \omega.\label{firstb}
\eeq
These equations imply that $\omega^{-1}F$ defines a transverse holomorphic structure that is integrable \cite{Kapustin:2001ij} giving us the $z_i$ coordinates. Given with $F$ satisfying \eqref{first} and \eqref{firstb},
we get a transverse holomorphic structure on $Y$ in which
\beq\label{THFc}
\hat{\Omega} =  F+i \, \omega,
\eeq
is a $(2,0)$ form.

\subsubsection{A Simple Example}
There is a simple way to satisfy all the conditions  if $v$ generates a symmetry of $X$ which not only preserves the symplectic form $\omega$, but also the holomorphic $(3,0)$ form $\Omega$, which we will use to construct explicit examples later.
In this case, the solution for $F$ can be taken to be
\beq\label{Fe}
F = {\rm Re}\, i_v\,\Omega.
\eeq
The idea here is that, if $v$ integrates to a group action on $X$, than a formal quotient of $Y$ by it has hyper-Kahler structure, whose $(1,1)$ form is the restriction of ${\omega}$, and whose $(2,0)$ form is $i_v\Omega$, which one can show is closed. Then \eqref{firstb} follows pointwise. Again, it is important to note the difference between the complex structure of THF, corresponding to ${\hat \Omega}$ in \eqref{THFc} being the holomorphic two-form, and that coming from $X$ in which $i_v\Omega$ is holomorphic. (For a completely explicit example, see section 6, which works out the case of $X=T^*S^3$ in detail.)
\subsection{Lagrangian 3-branes parallel to coisotropic 5-brane}\label{s:TFHt}

We now study the geometry of a parallel Lagrangian 3-brane and a coisotropic 5-brane in $X$, i.e., we have a coisotropic 5-brane $Y$ with field strength $F$ satisfying (\ref{firstb}) and a Lagrangian 3-brane $M \subset Y$ contained in $Y$. In many cases, we will take $X$ to be the cotangent bundle $T^*M$, and $Y$ to be a rank two bundle over $M$ which is a sub-vector bundle of $T^*M$. 

We saw above that the field strength $F$ of the coisotropic brane and the symplectic form $\omega$ define a THF on $Y$ by 
\beq\label{THFt}J = \omega^{-1} F.
\eeq
We can ask, when does this THF on $Y$ induce a THF on $M$? That is, when do there exist local coordinates $(z_1, z_2, t)$ on $Y$ such that $M$ is the set where $z_2 = 0$, so that $M$ is a holomorphic submanifold? By the integrability result of \cite{Kapustin:2001ij}, this occurs if $Ju$ remains tangent to $M$ for any vector $u$ tangent to $M$.

The answer is straightforward: the THF on $Y$ induces one on $M$ exactly in the case that 
$$F|_{M} = 0,$$ 
when the field strength on $Y$ vanishes when restricted to $M$. To see this, suppose first that $J= \omega^{-1} F$ preserves $TM$. Then, for vectors $u, w \in TM$, we have
\[ F(u, w) = \omega(u, J w) =0. \]
The first equality follows per definition of $J$, and the second since $Jw \in TM$ and $\omega_M=0$ since $M$ is Lagrangian. For the other direction, suppose $F|_M = 0$, and $u \in TM$. We must show that $J u \in TM$. Equivalently, if $g$ is any (locally defined) function on $X$ with $g|_M = 0$, we must show that $dg(J u) = 0$. But we have
\[ dg(Ju) = (\del_i g) \omega^{i j} F_{j k} u^k = v_g^j F_{j k} u^k = F(v_g, u) , \]
where $v_g$ is the Hamiltonian vector field induced by $g$. Since $M$ is Lagrangian, we have $v_g \in TM$, and because $F|_M = 0$, we have
\[ dg(Ju) = F(v_g, u) = 0, \]
as desired.

Finally, if we are given a 3-manifold $M$ with THF, we might look for a symplectic manifold $X$ and coisotropic 5-brane $Y$ such that the THF on $M$ is induced from $Y$ in the above manner. There is in fact a natural choice. We may set

\beq\label{setup}
X = T^*M, \qquad Y = T^*_{\perp} M,  \qquad F(-, -) = \omega(-, J-)|_Y 
\eeq
where $T^*_\perp M$ is the rank two sub-bundle consisting of cotangent vectors which are perpendicular to the foliation on $M$. In other words, if we choose local coordinates $(x, y, t)$ on $M$ such that $z = x + i y$ is holomorphic, with corresponding momenta coordinates $(p_x, p_y, p_t)$ for the fibers of the cotangent bundle, with $\omega = dt \wedge dp_t + dx \wedge dp_x+dy \wedge dp_y$.  Then, we have that $Y$ is the set where $p_t = 0$, and we have 
\[ F = - d p_x \wedge dy + d p_y \wedge dx. \]
It is easy to check that $F$ satisfies (\ref{firstb}), and thus $Y$ is a coisotropic A-brane in $T^*M$. In particular, this shows that even though our theory may be defined on any 3-manifold with THF without any reference to the topological string, it may always be embedded in the topological A-model regardless.

\subsubsection{Bi-fundamental Matter}
If we have $N$ Lagrangian 3-branes which are parallel to  $M$ coisotropic 5-branes, there is a new open string sector of 3-5 strings stretched between them. These give a matter structure which is charged in bifundamental representation under the gauge groups on the Lagrangian branes and the coisotropic branes. 
For the purposes of this paper we will take the 5-brane to be non-compact and non-dynamical. This gives $M$ fields transforming in the fundamental of $U(N)$.
Moreover, as we discussed, the 3-3 sector leads to CS theory of $U(N)$.  We thus end up with $U(N)$ Chern-Simons theory coupled to $M$ fields (and their conjugates) transforming in the fundamental of $U(N)$.   Further, as explained in the previous section when $F=0$ along the Lagrangian brane, the 3-5 system inherits a THF structure from the coisotropic brane.  Thus the part of the Lagrangian involving the interaction of the $U(N)$ gauge field with the fundamental matter field can only use the THF structure.  This essentially fixes the form of the action we wrote in the previous section (i.e. we cannot use any structure other than THF).

In the next section we will give a direct derivation of this from topological string field theory, in the special case when $X$ is a product manifold, $X=X_1\times X_2$ with $X_{1,2}$ Calabi-Yau manifolds of dimensions one and two.

\subsection{5d Chern-Simons Theory}\label{s:5da}
It is natural to ask what is the effective theory on the co-isotropic brane. We will propose two answers, which will show are equivalent classically. 

The first description is in terms of 5d Chern-Simons theory where an imaginary background gauge field has been turned on. The second is in terms of a non-commutative variant of this theory, which appeared recently in \cite{{Costello:2016}}. The two descriptions are related by Seiberg-Witten transform, as we will see. 

\subsubsection{5d ``Ordinary" Chern-Simons Theory}
Given that the theory on the Lagrangian brane is Chern-Simons theory in 3d, it is natural to propose that the 5d one is the 5d analog of it, i.e. the 5d Chern-Simons theory. However this cannot be exactly right for two reasons:  First, the 5d CS theory is not a well-defined quantum system because there is no quadratic action.  The smallest number of fields appearing is cubic.  Secondly, we know that topological string requires an $F$ compatible with $\omega$ for its consistency, whereas $A=0$ is a good solution to 5d CS theory, which violates this.  

These two problems are each other's solution:  Let $a$ be a 1-form where $da=\omega$.  Note that $a$ is only locally defined and in going from patch-to-patch it can shift by $a\rightarrow a+d\epsilon$.  In other words, we can think of $a$ as a connection whose field strength is $\omega$.
We propose the following Lagrangian for the 5d coisotropic brane:
$${1\over g_s}\int_{Y}\, {\rm CS}_5(A+i a)$$
Note that the theory starts with a quadratic term in $A$, of the form $AdA \wedge \omega$.  
This Lagrangian leads to the classical equations of motion:

$$(F+i\omega)\wedge (F+i\omega)=0\;\;\;  \rightarrow  \;\;\; F\wedge F=\omega \wedge \omega  \;\;  \textup{and} \;\;  F\wedge \omega=0,$$
which are exactly the equations \eqref{firstb} defining the coisotropic brane with the flux on it.  The classical equations of motion should lead to classical string solutions and this is indeed the case with this action.\footnote{It is natural to conjecture this theory is related to ${\cal N}=1$ supersymmetric theories  which are adapted to preserve supersymmetry on 5-manifolds.}

\subsubsection{5d Non-Commutative Chern-Simons Theory}
Another natural action, using the ingredients at hand, is the 5d non-commutative Chern-Simons theory. 

We consider $Y$ as a manifold with non-commutativity turned on, in direction given by the holomorphic two-form defined by a choice of THF. This means that the algebra of functions is 
\beq\label{MP}
f *_{\epsilon} g = f\, g + {\epsilon \over 2}\, {\hat {\Omega}}^{ij} {\partial \over \partial z^i} f \, {\partial \over \partial z^j} g +\ldots.
\eeq
where $\hat {\Omega}$ is defined in \eqref{THFc}. Non-commutativity introduces a scale, and the parameter $\epsilon$ helps us keep track of it; it accompanies every power of ${\hat \Omega}^{-1}$. The sub-leading terms in \eqref{MP} are defined so that the $*$ product is associative. For the most part, we will work in local coordinates in which ${\hat \Omega} = dz_1\wedge dz_2$ and then the $*_{\epsilon}$ product is just the Moyal product, see \eqref{MPr}. 

The action is
\beq\label{ncgen}
{1\over g_s \epsilon}  \int {\hat \Omega} \wedge ({1\over 2} {{\hat A}} *_{\epsilon} d{ {\hat A}} + {1\over 3} {{\hat A}} *_{\epsilon} {{\hat A}} *_{\epsilon} {{\hat A}}),
\eeq
where ${\hat A}$ is the non-commutative gauge-field, valued
 in
 $$
 {\hat A} \in \Omega^1(Y)/\Omega^{1,0}(Y) 
 $$
with modified gauge transformation properties that involve the $*_{\epsilon}$ product. Under infinitesimal $U(1)$ gauge transformation with parameter ${\hat \lambda}$, ${\hat A}$ transforms as,
$$
\delta_{\lambda} {\hat A} = d {\hat \lambda} + i {\hat \lambda} *_{\epsilon} {\hat A} - i  {\hat A}*_{\epsilon}{\hat  \lambda},
$$
which is a non-commutative deformation of the usual gauge symmetry.
 
In the next section, we will prove that, when $X$ is a suitable direct product manifold, the topological string field theory action is in fact given by \eqref{ncgen}. Namely, one needs $X = X_1\times X_2$, with $X_2$ a hyperkahler, and $Y = Y_1 \times X_2$ with $Y_1$ Lagrangian submanifold. Then, on $X_2$ we get the canonical coisotropic $A$-brane of \cite{Kapustin:2006pk}. It is natural to expect that this action is general, and holds for any $X$ with a coisotropic 5-brane on $Y$.

\subsubsection{Seiberg-Witten Transform}
The two actions are classically equivalent, related by Seiberg-Witten transform \cite{SW}. 

To explain this, first expand the commutative 5d Chern-Simons action around the solution given in \eqref{first},\eqref{firstb}, \eqref{THFc}. Writing the connection as the 
$$
A = A_0 + A', 
$$
where $A_0$ is the classical solution to \eqref{firstb}, which defines 
a THF complex structure on $Y$, in which
\beq\label{backgroundnc}{\hat \Omega} = F_0 +i \omega,
\eeq
becomes a holomorphic $(2,0)$ form, and 
$$
A' \in \Omega^1(Y)/\Omega^{1,0}(Y)
$$
we find
\beq\label{bckg}
{1\over 2 g_s} \int_Y\Bigl( {\hat \Omega} \wedge A' \wedge {\hat d}A'   + A' \wedge d A' \wedge {d}A' \Bigr).
\eeq
(The $(1,0)$ piece of $A'$ is deleted since it is not a physical field. This is evident in that \eqref{bckg} does not give it a kinetic term. This is an open string analogue of a similar phenomenon in the context of Kodaira-Spencer theory, see the discussion in \cite{BCOV}, p.  75.) 

Secondly, one should recall that in any theory with a D-brane, the $U(1)$ gauge field $F$ is always accompanied by a NS $B$-field and enters in combination
$$
F \;\; \rightarrow \; \; {\cal F} = F+B.
$$ As explained in \cite{wo}, this is necessary to preserve the $B$-field gauge invariance $B\rightarrow B+d\Lambda$ in presence of worldsheet boundaries on D-branes. To preserve this symmetry the $U(1)$ connection $A$ has to shift by $A \rightarrow A- \Lambda$. Correspondingly, all of our formulas so far should have the gauge field strength $F$ replaced by its invariant combination ${\cal F}$.
So far, we have been assuming that the background $B$ is zero. We could have equivalently used the $B$-field gauge transformation to set the background $F_0$ to zero, and turned on $B$ instead, keeping the combination $F_0+B$ fixed.
The freedom to trade $B$ for $F$ is referred to as the "$B$-field transform" in \cite{Kapustin:2005}.

In \cite{SW} Seiberg and Witten explained that a change of variables, at least classically, relates a non-commutative gauge theory to an ordinary one, while trading non-commutativity for $B$-field turned on. The derivation in \cite{SW} is general, independent of any details of the theory, and gives an in general non-linear relation between the non-commutative gauge fields ${\hat A}(\epsilon)$, and gauge parameters ${\hat \lambda}(\epsilon)$, at different values of non-commutativity parameter $\epsilon$, see eqn 3.8 of \cite{SW}. 

In our case, the Seiberg-Witten transform simply says one should identify the ordinary and the non-commutative gauge field 
$$A' = {\hat A},$$ 
and ${\hat \Omega}$ defined by the background in \eqref{backgroundnc}, with ${\hat {\Omega}}$ which enters the definition of the $*$-product in \eqref{MP}. With this, one easily shows that the actions \eqref{bckg} and \eqref{ncgen} are equal, up to total derivatives. For example, the second term in \eqref{bckg} comes from expanding the second term in \eqref{ncgen} to the first order in non-commutativity parameter. (For this, we used local coordinates $z_i$ in which ${\hat \Omega} = dz_1 \wedge dz_2$ is locally constant).

Since we can identify the non-commutative and ordinary gauge fields, we will drop the hats on ${\hat A}$ from now on.

\subsubsection{Comparison to \cite{Kapustin:2005}}
The relation between coisotropic A-branes with ordinary gauge invariance, and non-commutative gauge theories was noted first in \cite{Kapustin:2005}, in context of space-filing coisotropic branes on hyperkahler manifolds. Setting the curvature $F_0$ to zero on a coisotropic A-brane and turning on the $B$-field instead (the $B$-field transform) we end up with a space filling brane with no flux (and $B$-field turned on). Such a brane is naturally interpreted as a B-brane. The non-commutative gauge theory description comes from the Seiberg-Witten transform, relating a $B$-brane in a B-model with a B-field background, to a $B$-brane with non-commutativity turned on. 

One can bring this closer to our setting by taking a Calabi-Yau three-fold which is a product $X=X_a\times X_2$, with $X_2$ hyper-Kahler, and a coisotropic brane on $X$ which is space-filling on $X_2$. Then, the non-commutative gauge theory description arizes from a mixed A/B model, with A-model coming from $X_1$, and B-model from $X_2$.

In general, $X$ is not a direct product (for example, we will take $X=T^*S^3$ in section 6), so we cannot use this argument. It is possible that one would be able to define a mixed A/B-model for a general $X$ with a 2 dimensional integrable foliation, and a transverse holomorphic symplectic structure. One is likely to produce such models by using T-duality from a conventional B-model.  We will not attempt to do this here, and leave it as an interesting future project. In the above analysis, we were able to show the equivalence of \eqref{bckg} and \eqref{ncgen} purely at the level of field theory, and independently of the topological string origin of the non-commutative field theory in \eqref{ncgen}.


\section{Topological string field theory and Chern-Simons matter theory for $X=X_1\times X_2$}\label{s:tsft}

In this section, we study Calabi-Yau manifolds which are products $X = X_1 \times X_2$ of two lower dimensional Calabi-Yau manifolds, with $X_1$ of real dimension 2 and $X_2$ of dimension 4.
We will define and study topological string theory on $X$ which is a mixed  A/B-model, with A-model coming from $X_1$ and B-model from $X_2$. We will show that in this setting, we can reproduce rigorously the results from previous section. Namely, we will show that the theory on 3-branes in presence of 5-branes on is exactly the theory we defined in section 2. (Both the 3-branes and the 5-branes are taken to be Lagrangian $A$-branes on $X_1$, and holomorphic $B$-branes on $X_2$).

\subsection{Supersymmetries and topological twists}

In the product setting, when $X=X_1\times X_2$, the ${\cal N}=(2,2)$ $\sigma$-model with target $X$ acquires extra symmetries. The field content of this theory is simply the product of the field content of the model with target $X_1$ with the fields of the model with target $X_2$. Similarly, the Hilbert space of the theory is the tensor product of the Hilbert space of the model for $X_1$ with that for $X_2$.  From this we see that the symmetry algebras of the $\sigma$-model on $X_1$ and $X_2$ both act on the $\sigma$-model for $X_1 \times X_2$, and these actions commute with each other. 

This tells us that all of the symmetries we are familiar with in an ${\cal N}=(2,2)$ model appear twice in this situation.  For instance, there are $8$ instead of $4$ fermionic symmetries, $4$ coming from the supersymmetries on $X_1$ and $4$ from the supersymmetries in $X_2$.\footnote{One might imagine that having $8$ fermionic symmetries forces us to have $(4,4)$ supersymmetry instead of $(2,2)$. This clearly can not be the case, because a manifold with holonomy $SU(n) \times SU(m)$ is not generally hyper-K\"ahler, which is required by $(4,4)$ supersymmetry.  We are saved by the fact that the $8$ fermionic symmetries do not satisfy the commutation relations of the $(4,4)$ supersymmetry algebra. Four of the symmetries will commute to give the stress-energy tensor for the model on $X_1$, and the other four will give the stress-energy tensor for the model on $X_2$. The true stress-energy tensor of the model for $X_1 \times X_2$ is the sum of that for $X_1$ and $X_2$, and only a diagonal collection of $4$ supercharges will commute to give this.}

The ordinary ${\cal N}=(2,2)$ supersymmetry algebra is the diagonal subalgebra of this larger algebra. However, there is more than one way to find a copy of the ${\cal N}=(2,2)$ supersymmetry algebra in the larger algebra.   The $R$-symmetry of the ${\cal N}=(2,2)$ algebra is $O(2)_L \times O(2)_R$, a copy of $O(2)$ acting on the left and on the right moving supercharges. We will choose a reflection $\rho \in O(2)_R$ giving an automorphism of the ${\cal N}=(2,2)$ supersymmetry algebra. Then, instead of taking the diagonal subalgebra of the two copies of the $(2,2)$ algebra, we can take the algebra consisting of elements $(Q,\rho(Q))$, where $Q$ is a supersymmetry of the model on $X_1$ and $\rho(Q)$ is a supersymmetry of the model on $X_2$.  

In this way we find two \emph{different} copies of the $(2,2)$ supersymmetry algebra acting on the same theory, the supersymmetric $\sigma$-model with target $X_1 \times X_2$.  We are interested in performing a topological twist of the model where we use the non-standard action of the $(2,2)$ supersymmetry algebra.  Notice that the reflection $\rho \in O(2)_R$ interchanges the A-model and B-model topological twist. If we perform an A-twist of the model using this non-standard action of supersymmetry, it gives us a topological field theory which behaves as the A-model on $X_1$ and the B-model on $X_2$. 

 As usual, by integrating the correlators of this TFT over the moduli of surfaces, we can turn  this into a topological string theory.

\subsection{Topological string field theory at Chern-Simons level zero}
At level zero, we can find an exact match between the Chern-Simons matter theory we are considering and the  theory on a 3-brane in the presence of a 5-brane in a certain topological string theory. The topological string theory that is required to make this work is not a usual $A$- or B-model, but is instead a mixture of the two.

\subsubsection{Open string field theory in the mixed A-B model}
Supersymmetric boundary conditions for this model, with the non-standard action of the $(2,2)$ supersymmetry algebra, can arise as products of a supersymmetric boundary condition on $X_1$ with one on $X_2$. To be supersymmetric with respect to the supercharge which is the A-twist on $X_1$ and the B-twist on $X_2$, we should take the product of a special Lagrangian on $X_1$ with a coherent sheaf on $X_2$.  The cohomology, with respect to the sum of the supercharge and the BRST operator, of the space of states of the theory will be the tensor product of the cohomology on $X_1$ with the corresponding cohomology on $X_2$. This holds whether we consider the space of states on a circle, or on an interval with chosen boundary conditions.

Thus, if $L, L' \subset X_1$ are Lagrangians, and $\mc{F}, \mc{F}'$ are  coherent sheaves on $X_2$, the cohomology of the space of states where we take the boundary condition $L\times \mc{F}$ at $0$ and $L' \times \mc{F}'$ at $1$ is
\begin{equation}
HF^\ast(L,L') \otimes \op{Ext}^\ast(\mc{F},\mc{F'})
\end{equation}
where $HF^\ast$ is the Floer cohomology (calculating open-string states for branes on $X_1$) and $\op{Ext}^\ast$ is the Ext-groups for sheaves on $X_2$ (calculating open-string states for branes on $X_2$).

This will allow us to calculate the theory on a brane in the mixed $A/B$-model topological string theory.  Let us recall \cite{Witten:1992fb} the algorithm.  The open-string states for a string stretched between a brane and itself will form a dg algebra\footnote{Or, more generally, an $A_\infty$ algebra.} $\mc{A}$.  The differential is the BRST operator of the twisted theory.  This algebra will also have a trace, of cohomological degree (ghost number) $-3$.  The space of fields of the open-string field theory, in the BV-BRST formalism, is $\mc{A}[1]$. As before, $[1]$ refers to a shift in ghost number, so that elements of $\mc{A}^1$ are the physical fields of the theory, elements of $\mc{A}^0$ are ghosts for gauge symmetries, etc. The action functional is the Chern-Simons type functional
\begin{equation}
\tfrac{1}{2} \op{Tr} \alpha \d \alpha + \tfrac{1}{3} \op{Tr} \alpha^3.
\end{equation} 

Let us now specialize to the case when $X_1 = \R^2$ and $X_2 = \C^2$. We will use the discussion above to calculate the theory on a brane of the form $\R \times \C$, where $\R \subset \R^2$ is a Lagrangian $A$-brane and $\C$ gives a $B$-brane.  To describe the field theory, we first need to calculate the algebra of open-string states.  

The Floer cohomology of the Lagrangian $\R$ in $\R^2$ is, of course, just $\R$.  However, we would like to consider the Floer co-chains, rather than the Floer cohomology.  The Floer co-chains, in this case, can be modeled by the algebra $\Omega^\ast(\R)$ of differential forms on $\R$ (generally the differential and the algebra structure need to be corrected by holomorphic discs ending on a Lagrangian, but this can not happen here).

Next, let us calculate the open-string states on a $B$-brane $\C \subset \C^2$. It is a general result (due to Koszul) that if $V \to X$ is the total space of any vector bundle, then the Ext-algebra of the structure sheaf of $X$ in the total space of $V$ is
\begin{equation}
H^\ast_{\dbar}(X, \wedge^\ast V) 
\end{equation} 
that is, the Dolbeault cohomology of $X$ with coefficients in the exterior algebra of $V$. The grading is such that $\wedge^i V$ is in degree $i$. The natural co-chain model is $\Omega^{0,\ast}(X,\wedge^\ast V)$. 

Therefore the B-model open-string fields are $\Omega^{0,\ast}(\C)\otimes \wedge^\ast N$ where $N$ is the rank one normal bundle to $\C$ inside $\C^2$. The symmetries of the B-model on $\C^2$ must preserve the holomorphic symplectic form. It follows that the normal bundle to $\C$ inside $\C^2$ transforms under the $U(1)$ action as the cotangent bundle of $\C$, so that we can write the open-string fields as $\Omega^{0,\ast}(\C)[ \d z]$, where $\d z$ is of degree $1$.  

Combining with the open-string fields from the A-model factor, we find that the full space of open-string fields for a theory on $N$ branes wrapping $\R \times \C$ is\footnote{Strictly speaking we should use a \emph{completed} tensor product, which allows certain infinite sums.}
\begin{equation}
\Omega^\ast(\R) \otimes \Omega^{0,\ast}[\d z] \otimes \mf{gl}_N. 
\end{equation} 
This is the same as the space of differential forms $\Omega^\ast(\R \times \C)\otimes \mf{gl}_N$, but the differential is the operator 
\begin{equation}
\d_{\R} + \dbar_\C = \d t \,\partial_t + \d \zbar \,\partial_{\zbar}.
\end{equation}
By applying the algorithm mentioned above for constructing the field theory on the brane, we deduce that the field content in the BV formalism consists of 
\begin{align} 
\mc{B} &\in \d z \wedge  \Omega^{\ast}(\R \times \C) \otimes \mf{gl}_N \\ 
\mc{A}' & \in \Omega^\ast(\R \times \C) / (\d z \wedge \Omega^\ast(\R \times \C)) \otimes \mf{gl}_N[1]
\end{align} 
with action
\begin{equation}
\int \mc{B} \left( \d \mc{A}' + \tfrac{1}{2}[\mc{A}', \mc{A}'] \right).
\end{equation}
This is precisely the field content and action for the level $0$ limit of Chern-Simons theory discussed in section \ref{BV-BRST}.

\subsubsection{5-branes and bifundamental matter}
Let us introduce a 5-brane on $\R \times \C^2$ in $\R^2 \times \C^2$, which is a product of a Lagrangian $A$-brane on $\R$ and a space-filling $B$-brane on $\C^2$.  We will arrange the 3-brane so that its world-volume lies inside that of the 5-brane. 

Then the space of 5-3 strings can be computed, using reasoning similar to that presented above, to be
\begin{equation}
\Omega^\ast(\R) \otimes \Omega^{0,\ast}(\C).
\end{equation} 
The space of 3-5 strings is 
\begin{equation}
\Omega^\ast(\R) \otimes \Omega^{0,\ast}(\C) \d z [-1]
\end{equation} 
where $[-1]$ indicates  a shift in the grading. The reason for the presence of $\d z$ and this shift in the grading is that while $\op{Ext}^0(\Oo_{\C}, \Oo_{\C^2})$ is zero, $\op{Ext}^1(\Oo_\C, \Oo_{\C^2})$ consists of the space of holomorphic one-forms on $\C$.  Here $\Oo_\C$, $\Oo_{\C^2}$ are the structure sheaves of $\C \subset \C^2$ and of $\C^2$ respectively, and we recall that the space of open-string fields in the B-model is the Dolbeault resolution of the Ext-groups.

From this and from general considerations of open-string field theory we can find a description of the theory on $N$ 3-branes in the presence of $K$ 5-branes.  The field content is 
\begin{align} 
 \mc{B} &\in \d z \wedge  \Omega^{\ast}(\R \times \C) \otimes \mf{gl}_N \\ 
\mc{A}' & \in \Omega^\ast(\R \times \C) / (\d z \wedge \Omega^\ast(\R \times \C)) \otimes \mf{gl}_N[1]\\
\Phi & \in \d z \wedge \Omega^\ast(\R \times \C) \otimes \op{Hom} (\C^N, \C^K) \\
\Psi & \in \Omega^\ast(\R \times \C) / (\d z \wedge \Omega^\ast(\R \times \C)) \otimes \op{Hom}(\C^K, \C^N).
\end{align}
The fields $\Phi$ and $\Psi$ come from 3-5 and 5-3 strings respectively.  The action functional is 
\begin{equation}
\int \scB \d \A' + \tfrac{1}{2} \int \scB [\A',\A'] + \int \Psi \d \Phi + \int \Psi \A \Psi.
\end{equation}
This is precisely the field content and action functional for level zero Chern-Simons mattter theory in the BRST-BV formalism, as discussed in section \ref{BV-BRST}. 

\subsection{Topological string-field theory at general Chern-Simons level}
How can we modify this procedure to introduce the Chern-Simons level?  We have seen previously that we should be able to do this by using the A-model twist on $\C^2$ instead of the B-model twist. In this section we will instead derive the Chern-Simons matter theory at general level by making $\C^2$ non-commutative.  A result of Kapustin \cite{Kapustin:2004} tells us that these two procedures are equivalent. On any hyper-K\"ahler manifold, there a $\mbb{P}^1$ of topological field theories connecting the $A$ and B-models in a fixed complex structure.  If we take $0$ to represent the  B-model and $\infty$ to represent the A-model, then a generic point in this family of TFTs can be viewed either as a non-commutative B-model or as an A-model with a $B$-field. The parameter of non-commutativity is the inverse to the coefficient of the $B$-field. 

As the first step in our analysis, we will compute the open-string algebra for a 3-brane on $\R \times \C$ in $\R^2 \times \C^2$, where $\C^2$ is non-commutative.  Because the open-string algebra is a tensor product of the A-model and B-model open string algebras, and the A-model open-string algebra is simply $\Omega^\ast(\R)$, we only need to compute the B-model open-string algebra.  

To compute the Ext-groups of the sheaf $\Oo_\C$ on $\C^2$, we resolve this sheaf by vector bundles. The resolution is the two-term complex
\begin{equation}
\Oo_{\C^2} \xto{w} \Oo_{\C^2}
\end{equation} 
situated in degrees $-1,0$. We use coordinates $z,w$ and  where $\C \subset \C^2$ is the locus $w = 0$. 

The Ext-groups are the cohomology of the complex obtained by taking maps from this two-term complex to $\Oo_\C$.  Using the Dolbeault resolution, we find the Ext-groups are computed by
\begin{equation}
\Omega^{0,\ast}(\C) \xto{w} \Omega^{1,\ast}(\C)
\end{equation}
where $\Omega^{1,\ast}(\C)$ is situated in degree $1$.

If we were considering the ordinary commutative B-model, then multiplication by $w$ acts by zero on $\Omega^{0,\ast}(\C)$.  In the non-commutative B-model, however, the operator $w$ acts by $\eps \partial_z$ where $\eps$ is the parameter of non-commutativity. It follows that the B-model open-string states give the algebra $\Omega^\ast(\C)$ with differential $\eps \partial + \dbar$.  

Including the A-model open string states on $\R$ gives $\Omega^\ast(\R \times \C)$ with differentail $\d_{\R} + \eps \partial_\C + \dbar_\C$.  This tells us that the open-string field theory on a stack of $N$ 3-branes has fields, in the BV-BRST formalism,
\begin{align}
\mc{A} &\in \Omega^\ast(\R \times \C) \otimes \mf{gl}_N[1] \\
\lambda S(\mc{A}) & = \tfrac{1}{2}\int \op{Tr}  \left( \mc{A} \d_{\R} \mc{A} + \mc{A} \dbar_\C\mc{A} + \eps \mc{A} \partial_\C \mc{A} \right) + \tfrac{1}{3} \int \op{Tr} \mc{A}^3.  
\end{align} 
Here we have introduced a coupling constant $\lambda$.  If we perform a change of coordinates which multiplies the components of $\mc{A}$ which involve $\d z$ by $\eps^{-1}$, the action becomes
\begin{equation}
S(\mc{A}) = \frac{1}{g_s} \int CS(\mc{A}),
\end{equation} 
where the topological string coupling constant $g_s$ is related to $\lambda$ and $\epsilon$ as $\epsilon/\lambda = 1/g_s$. 
(Recall that the Chern-Simons level $k$ is related to the topological string coupling by $g_s = {4\pi i \over k+N}$, taking into account the one loop shift.)

\subsubsection{Bifundamental matter}
Let us introduce a 5-brane on $\R \times \C^2$ which is parallel to the 3-brane.  The analysis of 5-3 and 3-5 strings is identical to that explained in section \ref{BV-BRST} in the case of level zero. The space of 5-3 strings is
\begin{equation}
\Omega^\ast(\R) \otimes \Omega^{0,\ast}(\C) 
\end{equation}
and the space of 3-5 strings is
\begin{equation}
\Omega^\ast(\R) \otimes \Omega^{1,\ast}(\C)[-1].
\end{equation}
The theory on $N$ 3-branes in the presence of $K$ 5-branes has fields  
 \begin{align} 
\mc{A} & \in \Omega^\ast(\R \times \C) \otimes \mf{gl}_N[1]\\
\Phi & \in \d z \wedge \Omega^\ast(\R \times \C) \otimes \op{Hom} (\C^N, \C^K) \\
\Psi & \in \Omega^\ast(\R \times \C) / (\d z \wedge \Omega^\ast(\R \times \C)) \otimes \op{Hom}(\C^K, \C^N).
\end{align}
The fields $\Phi$ and $\Psi$ come from 3-5 and 5-3 strings respectively.  The action functional is 
\begin{equation}
\frac{1}{g_s} \int CS(\A)  + \int \Psi \d_{\A} \Phi 
\end{equation}
This is the field content and action functional for the Chern-Simons mattter theory in the BRST-BV formalism, as discussed in section \ref{BV-BRST}. 

\subsubsection{Comparing with the A-model picture}\label{s:rot}
Let us discuss in more detail how this picture compares with the A-model picture discussed earlier.  For any two-dimensional $(4,4)$ theory, there is a $\mbb{P}^1 \times \mbb{P}^1$ of $(2,2)$ subalgebras in the $(4,4)$ algebra.  In the case of the $\sigma$-model with target a hyper-K\"ahler manifold $X$, the a $(2,2)$ subalgebra is given by a pair $(J_1,J_2)$ of complex structures on $X$ \cite{Kapustin:2006pk}.  We will fix $J_1 = J$, and vary $J_2$.  We will also fix once and for all one way of topologically twisting a theory with $(2,2)$ - say the B-twist.  

The topological twist for the $(2,2)$ subalgebra corresponding to $(J,J)$ is the B-model on $X$ in complex structure $J$. The topological twist for the subalgebra corresponding to $(J,-J)$ is the A-model in complex structure $J$.  As explained in \cite{Kapustin:2004, Kapustin:2005}, the topological twist in a general complex structure $(J,J')$ can be interpreted either as a non-commutative B-model or as an A-model with a $B$-field.  

If we have a product $X_1 \times X_2$, where $X_1$ is K\"ahler and $X_2$ is hyper-K\"ahler, then we have a family of ways of equipping the supersymmetric $\sigma$-model with an action of the $(2,2)$ supersymmetry algebra, parameterized by the choice of $(2,2)$ algebra inside the $(4,4)$ algebra acting on the theory on $X_2$. By varying this choice, and performing a topological twist, we find a $\mbb{P}^1$ of topological field theories interpolating between the A-model on $X_1 \times X_2$ and the mixed A/B-model on $X_1 \times X_2$. The generic point in this family can be interpreted as either an A-model on $X_1 \times X_2$ with a $B$-field coming from a $2$-form on $X_2$, or as a mixed A/B-model where the B-model in $X_2$ is made non-commutative.  

When there is a non-zero $B$-field on $X_2$, we can introduce a coisotropic brane that wraps $X_2$ and a Lagrangian in $X_1$.  In the mixed A/B-model interpretation of the same theory, this coisotropic brane becomes a product of a Lagrangian in $X_1$ and a space-filling $B$-brane in the non-commutative B-model on $X_2$. 

This explains why the purely A-model realization of the Chern-Simons matter theory should be equivalent to the realization in the mixed A/B-model.  

\subsubsection{The theory on a 5-brane}
Let us return to studying the mixed A-B model on $\R^2 \times \C^2$, where $\C^2$ is made non-commutative with parameter of non-commutativity $\eps$.  We have seen that the Chern-Simons matter theory, at level $1/g_s =\eps/\lambda$ is realized as the theory on a 3-brane in the presence of a parallel 5-brane.

The question what is the theory on a 5-brane is easily answered by a topological string field theory analysis similar to the one given earlier.  The open-string states for a string stretched between two stacks of $N$ 5-branes is

\begin{equation}
\Omega^\ast(\R) \otimes \Omega^{0,\ast}(\C^2) \otimes \mf{gl}_N = \Omega^\ast(\R \times \C^2) / ( \d z_1 \wedge \Omega^\ast, \d z_2 \wedge \Omega^\ast ) \otimes \mf{gl}_N.
\end{equation}
The differential is

\begin{equation}
\d_\R + \dbar_{\C^2} = \d t \partial_t + \d \zbar_1 \partial_{\zbar_1} + \d \zbar_2 \partial_{\zbar_2}.
\end{equation}
When $\eps = 0$, the algebra structure is simply derived from the wedge product of forms.  However, if we turn on the non-commutativity, the algebra structure is deformed by the Moyal product.  If $\alpha,\beta$ are open-string fields, the Moyal product is
\begin{multline}\label{MPr}
 \alpha \ast_\eps \beta = \alpha \wedge \beta + \eps  \tfrac{1}{2} \op{Alt}_{ij} \dpa{z_i}\alpha  \wedge \dpa{z_j} \beta \\
  + \eps^2 \tfrac{1}{2^2 \cdot 2!} \op{Alt}_{i_1 j_1} \op{Alt}_{i_2 j_2} \left( \dpa{z_{i_1}} \dpa{z_{i_2}} \alpha \right) \wedge \left( \dpa{z_{j_1}} \dpa{z_{j_2}} \beta \right) + \dots
\end{multline} 
From this expression for the product we can derive the action functional for the theory living on the brane. The fields, in the BV-BRST formalism, are
\begin{equation}
\mc{A} \in \Omega^\ast(\R) \otimes \Omega^{0,\ast}(\C^2) \otimes \mf{gl}_N [1]. 
\end{equation}
Thus, the fundamental field, of ghost number zero, is a three-component partial connection
\begin{equation}
A = A_t \d t + A_{\zbar_1} \d \zbar_1 + A_{\zbar_2} \d \zbar_2.
\end{equation}
The other fields are ghosts for the natural gauge symmetry of this connection, anti-fields, and anti-ghosts.  

The action functional is 
\begin{equation}
\int \d z_1 \d z_2\tfrac{1}{2} \op{Tr} (\mc{A} \d \mc{A}) + \int \d z_1 \d z_2 \tfrac{1}{3}\int  \op{Tr} (\mc{A} \ast_\eps \mc{A} \ast_\eps \mc{A} ). 
\end{equation}
Thus, we find the theory is a 5-dimensional non-commutative Chern-Simons theory.

In fact, exactly this theory was studied in great detail in \cite{Costello:2016}, where it was argued that this theory captures the super-symmetric part of $11$-dimensional supergravity in an $\Omega$-background. It would be nice to find a physical link between the M-theory setup in \cite{Costello:2016} and the one here.

 The derivation presented here, using the mixed A/B-model, applies for direct product manifolds $X=X_1\times X_2$. For the reasons explained in section \ref{s:5da} we believe the result holds more generally, for any coisotropic 5-brane in a non-compact Calabi-Yau $X$.

\section{A direct link with $3d$ ${\cal N}=2$ theories}
We have argued that the theories we are considering are related to 3-dimensional ${\cal N}=2$ theories. In this section we will show directly that a partially topological twist of the 3-dimensional ${\cal N}=2$ gauge theory gives rise to the theories we are considering.  

\subsection{Generalities on twisting}
We should first discuss what we mean by twisting a $3d$ theory with ${\cal N}=2$ supersymmetry.  
 The $R$-symmetry group for a $3d$ ${\cal N}=2$ theory is $U(1)$.   It is not possible to find a homomorphism from $\op{Spin}(3)$ to $U(1)$ with which we can change the spin of the fields.  Therefore, there is no topological twist of $3d$ ${\cal N}=2$ theory in the traditional sense introduced by Witten.

Instead we will consider a twist which is invariant under $\op{Spin}(2)$ instead of $\op{Spin}(3)$. To describe how such twists behave, let us recall the structure of the ${\cal N}=2$ supersymmetry algebra. There are $4$ supercharges which have charge $(\pm \tfrac{1}{2}, \pm 1)$ under the action of $\op{Spin}(2)$ and $U(1)_R$. We let $Q_{\pm \tfrac{1}{2}, \pm 1}$ be the supersymmetry charged in this way. We have the commutation relations
\begin{align}
[Q_{\tfrac{1}{2}, 1}, Q_{\tfrac{1}{2}, -1}]  &=\partial_{z}\\
[Q_{-\tfrac{1}{2}, 1}, Q_{-\tfrac{1}{2}, -1}]  &=\partial_{\zbar}\\
[Q_{\tfrac{1}{2}, 1}, Q_{-\tfrac{1}{2}, -1}]  &= \partial_t
\end{align}
where we have chosen coordinates $t,z$ on $\R^3 = \R \times \C$, in which the line with coordinate $t$ is invariant under $\op{Spin}(2) \subset \op{Spin}(3)$. 

If we twist using the identity homomorphism $\op{Spin}(2) \to U(1)_R$, we find that $Q_{\tfrac{1}{2}, 1}$ and $Q_{-\tfrac{1}{2}, -1}$ transform as scalars under the twisted action of $\op{Spin}(2)$.  We choose the supercharge $Q = Q_{-\tfrac{1}{2}, -1}$ to be the one we twist with. This means that we add this supercharge to the BRST operator of the physical theory.  

After twisting, the operators $\partial_t$ and $\partial_{\zbar}$ are BRST exact. They commute with the physical BRST operator, but are exact for the supersymmetry $Q$, and so are exact for the twisted BRST operator
\begin{equation}
Q_{BRST}^{twisted} = Q_{BRST}^{physical}  + Q. 
\end{equation}
This implies that correlators of local operators which are closed for $Q_{BRST}^{twisted}$ are functions of the positions of the operators which are independent of $t$ and of $\zbar$. The same therefore holds for the OPE.  In addition the OPE can't have any singularities, because it is independent of $t$ and does not have singularities if $t \neq 0$.  

How do we globalize this local story? Clearly everything we have described so far works for a spin 3-manifold whose holonomy is $SO(2) \subset SO(3)$. Such a 3-manifold has a transverse holomorphic foliation structure.  Indeed, the subbundle of the tangent bundle which locally is preserved by the $SO(2)$ inside $SO(3)$ is integrable, since all rank one bundles are integrable.  The leaf space has a metric of $SO(2)$ holonomy and so a complex structure. 

However, not all manifolds with a THF structure have a metric with $SO(2)$ holonomy.  We would like to be able to put a $3d$ ${\cal N}=2$ theory on any 3-manifold with a THF structure in such a way that the supercharge $Q$ is a symmetry of the theory.  

Let us consider putting $3d$ ${\cal N}=2$ theories on general ${\cal N}=2$ supergravity backgrounds, where we use the version of ${\cal N}=2$ supergravity for which the $R$-symmetry is gauged.  We can twist any $3d$ ${\cal N}=2$ theory placed on a  background in which there is a generalized Killing spinor of square zero.  3-manifolds of holonomy $SO(2)$ provide examples of such backgrounds: if we ask that the gauge field for the $R$-symmetry is the Levi-Civita connection (viewed as an $SO(2)$ connection), then there are two generalized Killing spinors.  The spinors are flat for the connection which is a sum of that coming from the metric and the $R$-symmetry background gauge field. (Of course, this is just the global version of the local calculation we described above). 

Because $3d$ ${\cal N}=2$ supergravity has more bosonic fields than just a metric and an $R$-symmetry gauge field, we can imagine trying to construct more general supersymmetric backgrounds than this.  In \cite{Klare:2012gn, Closset:2012ru, Closset:2013vra}  it was shown that for any 3-manifold with a transverse holomorphic structure there exists a supergravity background with a generalized Killing spinor $\Psi$ satisfying $\Gamma(\Psi,\Psi) = 0$.  Thus, any ${\cal N}=2$ theory can be twisted when it is placed on such a background.

\subsection{Twisting the vector multiplet}
 In this section we will calculate the twist of the ${\cal N}=2$  vector and chiral multiplets, and show that the twisted theories are the ones we wrote earlier.  This will give a description of the twisted $3d$ ${\cal N}=2$ theory on any manifold with a THF structure in a way which manifestly only depends on the transverse holomorphic fibration.  
We will calculate the twists explicitly in coordinates on flat space. Globalization is straightforward.

First, we will show that a twist of $3d$ ${\cal N}=2$ pure gauge theory, with a level $k$ Chern-Simons term, is equivalent to 3-dimensional Chern-Simons theory at level $k$. Without the Chern-Simons term we find the level $0$ limit of Chern-Simons that we discussed in section \ref{BV-BRST}. 

In fact, this result will follow from a calculation performed in \cite{Costello:2013zra}. There the holomorphic twist of $4d$ ${\cal N}=1$ pure gauge theory was calculated. It was shown that the holomorphic twist is equivalent to holomorphic BF theory on four dimensions, whose field content consists of 
\begin{align}
A^{0,1} &\in \Omega^{0,1}(\C^2) \otimes \mf{g} \\
B^{2,0} & \in \Omega^{2,0}(\C^2) \otimes \mf{g}.
\end{align} 
The action functional is 
\begin{equation}
\int B F^{0,2}(A).
\end{equation}
The infintesimal gauge symmetries are
\begin{align}
A^{0,1} & \mapsto \dbar \eps + [\eps, A]\\
B^{2,0} & \mapsto [\eps, B^{2,0}] 
\end{align}
where $\eps \in \Omega^{0,0}(\C^2) \otimes \mf{g}$ is the generator of infinitesimal gauge transformations.

Let us reduce this theory to three dimensions.  We will choose coordinates $z,w$ and write $w = t + i x$. We'll reduce along the $x$ coordinate. Since $A^{0,1}$ has only two components, it gives rise to a two component connection in three dimensions,
\begin{equation}
A^{3d} = A^{3d}_t \d t + A^{3d}_\zbar \d \zbar
\end{equation}
where $A^{3d}_t = A^{0,1}_{\br{w}}$ and $A^{3d}_\zbar = A^{0,1}_{\zbar}$.  The field $B^{2,0}$ reduces to a single-component field $B^{3d}$ which transforms under $SO(2)$ as a $(1,0)$ form $f(t,z,\zbar) \d z$. 

The action functional in three dimensions is 
\begin{equation}
\int B^{3d} F(A^{3d}).
\end{equation}
This is the level $0$ limit of Chern-Simons theory discussed in section \ref{BV-BRST}.

\subsubsection{Relating to the fields of the physical theory}
Let us explain the origin of these fields in the physical theory, from the four-dimensional point of view.  We can write four-dimensionl ${\cal N}=1$ gauge theory in the first-order formulation, where the fundamental fields are
\begin{align}
A & \in \Omega^1(\R^4) \otimes \mf{g} \\
B & \in \Omega^2_+ (\R^4) \otimes \mf{g}\\
\Psi_{\pm} & \in \cinfty(\R^4, S_{\pm}) \otimes \mf{g} 
\end{align}
so that $A$ is an ordinary gauge field, $B$ is an adjoint-valued self-dual $2$-form, $\Psi_{\pm}$ are sections of the two rank two spinor\footnote{Because we work in Euclidean signature, we must treat the spinor fields as complex fields, and perform the path integral over a contour.  This is because the four-dimensional spin representation $S_+ \oplus S_-$ of $\op{Spin}(4,\C)$ does not admit a real structure.} bundles $S_{\pm}$. The action functional is 
\begin{equation}
\int B \wedge F(A) + c \int B \wedge B + \int \Psi_+ \slashed{\partial}_A \Psi_.
\end{equation}
The coupling constant $c$ is related to the usual Yang-Mills coupling constant by a simple transformation. 

In \cite{Costello:2013zra} it is shown that variation of $c$ is $Q$-exact for the supercharge we use to twist, so that we can set $c = 0$. (We choose a supercharge in $S_-$,which breaks Lorentz symmetry to $SU(2)_+$ and allows us to decompose differential forms into their $(p,q)$ types).   It is then shown that the action of supersymmetry cancels the $(1,0)$ components of the gauge field $A$ with the spinors $\Psi_-$, and cancels two of the three components of $B$ with $\Psi_+$.  We are left with the $(2,0)$ component of $B$ and the $(0,1)$ component of $A$, and the holomorphic BF action. 

\subsubsection{Adding a Chern-Simons term}
In \cite{Costello:2013zra} the effect of adding a certain four-dimensional Chern-Simons term to ${\cal N}=1$ gauge theory was considered, leading to a four-dimensional cousin of Chern-Simons. Here we will see that adding an ordinary three-dimensional Chern-Simons term to $3d$ ${\cal N}=2$ theory will lead, after twisting, to pure Chern-Simons gauge theory.

The four-dimensional Chern-Simons term considered in \cite{Costello:2013zra} is 
\begin{equation}
\int \d w CS(A) + \int \psi_+ (\d w \cdot \psi_-)
\end{equation}  where $\d w \cdot$ indicates Clifford multiplication.  In \cite{Costello:2013zra}, it is shown that, if we add this Chern-Simons term in $4$-dimensions and then twist using the supercharge $Q \in S_-$, we add the term
\begin{equation}
\int A^{0,1} \partial_z A^{0,1} \d z \d w
\end{equation}
to the action for holomorphic BF theory (the gauge symmetry for the field $B^{2,0}$ is also changed). Reducing to three dimensions, we find that we add the term 
\begin{equation}
\int A^{3d} \partial_z A^{3d} \d z 
\end{equation} 
to the action for level $0$ Chern-Simons. As discussed in section \ref{BV-BRST} this gives us Chern-Simons at non-zero level.

\subsection{Twisting theories with matter}
Let's now analyze what happens when we perform the twist of a gauge theory with matter. We will calculate the twist in $4$-dimensions, and then reduce to 3-dimensions.  To start with, we will calculate the twist of the free $4d$ ${\cal N}=1$ chiral multiplet.

Recall that we can arrange the fields of the $4d$ ${\cal N}=1$ chiral multiplet into a chiral superfield (and its complex conjugate).  Let us introduce two odd complex variables $\theta_\alpha$ which transform in the representation $S_+$ of $\op{Spin}(4)$, and two odd complex variables $\theta_{\dot{\alpha}}$ transforming in $S_-$.  Then, the complexified space of fields of the $4d$ ${\cal N}=1$ chiral multiplet consists of 
\begin{align}
\Phi_+ & \in \cinfty(\R^4) [\theta_{\alpha}]\\
\Phi_- & \in \cinfty(\R^4)[\theta_{\dot{\alpha}}] . 
\end{align}
We can expand each field into components
\begin{align} 
\Phi_+ &= \phi_+ + \theta_{\alpha} \psi_+^{\alpha} + \theta_{\alpha} \theta_{\beta} \eps_{\alpha \beta} F_+ \\
 \Phi_- &= \phi_- + \theta_{\dot{\alpha}} \psi_-^{\dot{\alpha}} + \theta_{\dot{\alpha}} \theta_{\dot{\beta}} \eps_{\dot{\alpha} \dot{\beta} } F_-. 
\end{align}
The fields $\phi_{\pm}$ are the chiral scalar and its complex conjugate, $\psi_{\pm}$ are the spinors of the chiral multiplet, and $F_{\pm}$ are auxiliary fields.  This is the complexified space of fields, so that the functional integral is taken over a contour.  The contour chosen for the fermionic fields does not matter, because small perturbations of the contour do not change the result and the fermions can be treated perturbatively. The contour for the bosonic fields does matter: one should take $\phi_+ = \overline{\phi_-}$, $F_+ = \overline{F_-}$. (One point that sometimes generates confusion is that there is no sense in which $\psi_-$ is the complex conjugate of $\psi_+$, because the complex conjugate of the representation $S_+$ of $\op{Spin}(4)$ is $S_+$ not $S_-$.) 

The action functional is the sum of a kinetic term and a superpotential term. We will discuss the superpotential term momentarily. The kinetic term is
\begin{equation}
\int \partial_{x_i} \phi_+ \partial_{x_i} \phi_- + \int \psi_+ \slashed{\partial} \psi_- + \int F_+ F_-. 
\end{equation} 

The fermionic elements of the ${\cal N}=1$ supersymmetry algebra act on the these fields by the operators
\begin{align}
D^\alpha &=  \partial_{\theta_\alpha} +  \Gamma^{\alpha \dot{\beta}}_i \theta_{\dot{\beta}} \partial_{x_i} \\ 
D^{\dot{\alpha}} &=  \partial_{\theta_{\dot{\alpha}}} +  \Gamma^{\beta \dot{\alpha}}_i \theta_{\beta} \partial_{x_i}  
\end{align}

The $U(1)$ $R$-symmetry group gives the variables $\theta_\alpha$ $R$-charge $+1$ and the variables $\dot{\theta}_{\dot{\alpha}}$ $R$-charge $-1$.  Thus, the fields $\psi_{\pm}$ have $R$-charge $\pm 1$ and $F_{\pm}$ have $R$-charge $\pm 2$.  

We twist this theory by choosing a supercharge 
\begin{equation}
Q = D^{\dot{1}} 
\end{equation}
and adding it to the BRST operator of the theory. Note that the choice of $Q$ breaks the Lorenz group to $SU(2)_+$. We can give the twisted theory an action of $SU(2)_+ \times U(1)$ by saying that the extra $U(1)$ acts by embedding it as the diagonal in $U(1)_R$ and the Cartan of $SU(2)_-$. With this choice the supercharge $Q$ is preserved by the extra $U(1)$. 

Let us see what happens to the chiral and anti-chiral multiplets when we twist in this way.  The operator $Q$ acts on the anti-chiral multiplet by the vector field $\partial_{\theta_{\dot{1}}}$.  This operator cancels the fields in the anti-chiral multiplet in pairs: one component of $\psi_-$ cancels with $\phi_-$, and the other cancels with $F_-$.

Next, let's examine the chiral multiplet. The action of $SU(2)_+ \times U(1)$ on the chiral multiplet factors through an action of $U(2)$.  The field $\phi_+$ is a scalar under $U(2)$, the spinors in $\psi_+$ transform under the fundamental representation of $U(2)$, and $F_+$ transforms in the exterior square of the fundamental representation.  Further, the action of $U(2)$ on $\R^4$ is by the usual embedding of $U(2)$ into $SO(4)$. 

It follows that we can identify the fields in the chiral multiplet with the space $\Omega^{0,\ast}(\C^2)$, where we view $\phi_+$ as an element of $\Omega^{0,0}(\C^2)$, $\psi_+$ as an element of $\Omega^{0,1}(\C^2)$, and $F_+$ as an element of $\Omega^{0,2}(\C^2)$. 

It is natural to give the fields of the theory a cohomological grading by their $R$-charge. In this way $\Omega^{0,i}(\C^2)$ is placed in degree $i$.  

The operator $Q$ acts on the chiral multiplet by 
\begin{equation}
\theta_\alpha \Gamma^{\alpha \dot{1}} \partial_{x_i}.
\end{equation}
Under the identification of our fields with the Dolbeault complex, the operator $Q$ becomes the Dolbeault operator $\dbar$.

Finally, let us find what happens to the action.  The kinetic term in the action depends linearly on both  the chiral and anti-chiral multiplets. Since the fields in the anti-chiral multiplet are cancelled in pairs by supercharge $Q$, and the action is $Q$-invariant, it must also be $Q$-exact. (The superpotential term in the action is not $Q$-exact, and will be analyzed shortly). 

To sum up, we have found that after twisting, the action functional is $Q$-exact, the anti-chiral multiplet has no $Q$-cohomology, and that the chiral multiplet, with the action of $Q$, can be identified with the Dolbeault complex.

We would like to use this information to write down a Lagrangian description of the twisted theory.  What we need to do is to construct an action functional whose BRST operator will realize the operator $Q$ on the chiral multiplet. After all, when we twist we should add the chosen supersymmetry to the BRST operator, and since  the original  action functional is $Q$-exact, this is all their will be.

The BV formalism provides a straightforward mechanism to achieve this. We introduce anti-fields corresponding to the fields in the chiral multiplet. (Since the anti-chiral multiplet cancels, we need not consider it further).  The anti-fields to the chiral multiplet consist of 
\begin{align}
\phi_+^\ast & \in \Omega^{2,2} (\C^2)  \\
\psi_+^\ast & \in \Omega^{2,1}(\C^2) \\
F_+^\ast & \in \Omega^{2,0}(\C^2).
\end{align}
If, as above, we give the fields of the chiral multiplet a cohomological degree corresponding to their $R$-charge, then the fields of the anti-chiral multiplet must be given a cohomological degree so that the BV odd symplectic pairing is of degree $-1$.  With this grading, the field $\phi_+^\ast$ has degree $1$, $\psi_+^\ast$ has degree $0$, and $F_+^\ast$ has degree $-1$.  

Let us summarize what we have found so far. Consider  the twist of the free ${\cal N}=1$ chiral multiplet in $4$ dimensions valued in a complex vector space $V$.   We thus find that, after twisting and cancelling the anti-chiral multiplets, the field content of the ${\cal N}=1$ chiral multiplet in the BV formalism consists of 
\begin{equation}
\Omega^{0,\ast}(\C^2) \otimes V \oplus \Omega^{2,\ast}(\C^2) \otimes V^\ast[1] 
\end{equation}
where the symbol $[1]$ means that cohomological degrees are shifted down.  The BRST operator, acting on the space of fields, is the $\dbar$ operator.  The BV action, whose Hamiltonian vector field is the BRST operator, is
\begin{equation}
\int \alpha \dbar \beta
\end{equation}
where we let $\alpha$ denote the field in $\Omega^{0,\ast}(\C^2) \otimes V$ and $\beta$ the field in $\Omega^{2,\ast}(\C^2) \otimes V^\ast[1]$. 

\subsection{Generalization to curved spaces}
This result can be easily generalized. We can replace $V$ by an arbitrary K\"ahler manifold $Z$ and $\C^2$ by an arbitrary surface $X$ with $SU(2)$ holonomy.  Then we find that the twist of the ${\cal N}=1$ $\sigma$-model in $4$ dimensions has fields consisting of 
\begin{align}
\alpha& : \br{T}[1] X  \mapsto Z \\
 \beta & \in \Gamma(\br{T}[1] X, K_X \otimes \alpha^\ast T^\ast Z)
\end{align}
with action functional
\begin{equation}
\int_{T[1] X} \beta \dbar \alpha. 
\end{equation}
Here $\br{T}[1] X$ denotes the complex supermanifold whose fibres are the $(0,1)$ tangent bundle on $X$, placed in degree $-1$. We treat this as a supermanifold where we consider holomorphic functions only in the odd directions but smooth complex-valued functions in the even directions. Functions on this supermanifold (when equipped with a natural differential) are the Dolbeault complex on $X$.  Also $T^\ast Z$ indicates the $(1,0)$ cotangent bundle of $Z$.

(We have described the twisted theory with a complex space of fields and a holomorphic action functional. In perturbation theory this is a sufficient description, but non-perturbatively one has to specify  a contour).

\subsection{The superpotential}
Let us now return to analyzing the superpotential term in the action. Fix a holomorphic function $W$ on a complex vector space $V$.  As before, let
\begin{equation}
\Phi_+ \in \cinfty(\R^4)[\theta_\alpha] \otimes V
\end{equation}
denote the super-field of the chiral multiplet valued in $V$. We can view $\Phi_+$ as a map of supermanifolds
\begin{equation}
\Phi_+ :  \R^4 \times \C^{0 \mid 2} \to V
\end{equation}
(again viewing $\C^{0\mid 2}$ as a supermanifold where we only consider holomorphic, and not anti-holomorphic, functions).  The superpotential term in the action for the ${\cal N}=1$ chiral multiplet is of the form
\begin{equation}
\int \d \theta_1 \d \theta_2 W (\Phi(x,\theta_1,\theta_2)) . 
\end{equation}
It is immediate from the above analysis that, after twisting, the superpotential term contributes a function of the field $\alpha \in \Omega^{0,\ast}(\C^2,V)$ of the form
\begin{equation}\label{supp}
\int \d z_1 \d z_2 W(\alpha).
\end{equation}
We can write this more explicitly in components.  Choose a basis of $V$ and write the components of $\alpha$ as $\alpha_{i,0}$, $\alpha_{i,\zbar_1}$, $\alpha_{i, \zbar_2}$, $\alpha_{i,\zbar_1 \zbar_2}$.  We also write $\alpha_0 : \R^4 \to V$ to mean the total degree $0$ component of $\alpha$. Then, the functional is given by this functional is given by
\begin{equation}
\int \d z_1 \d z_2 \d \zbar_1 \d \zbar_2  \left( \partial_i W (\alpha_{0}) \alpha_{i,\zbar_1 \zbar_2} + \partial_i \partial_j W(\alpha_0) \alpha_{i,\zbar_1} \alpha_{j,\zbar_2} \right).  
\end{equation}

Note that the introduction of the super-potential term breaks certain symmetries we had before. The theory is no longer $U(2)$ invariant, it is only $SU(2)$ invariant, because the volume form $\d z_1 \d z_2$ appears in the superpotential term in the action. The theory is also only $\Z/2$, and not $\Z$, graded.  

We can rectify at least the first problem by assuming that the superpotential is quasi-homogeneous of weight $1$ for some grading on $V$.  If we make this assumption, then $V$ has a decomposition into weight spaces $V = \oplus V_\lambda$ where each weight space has weight a non-negative rational number $r(\lambda) \in \Q$.  We can then modify our space of fields by saying that 
\begin{align}
\alpha & \in \oplus_\lambda \Omega^{0, \ast} (\C^2, V_\lambda \otimes K^{r(\lambda)}) \\
\beta & \in \oplus_\lambda \Omega^{2, \ast} (\C^2, V^\ast_\lambda \otimes K^{-r(\lambda)})[1].
\end{align}
Then, we can define the action functional of the theory without using the holomorphic volume form on $\C^2$, and the action makes sense on any complex surface.
\subsection{Reduction to three dimensions}
Now that we have calculated the twist of the four-dimensional ${\cal N}=1$ chiral multiplet, we can derive a description of the twist of the three-dimensional ${\cal N}=1$ chiral multiplet.   If we work on $\C^2$ with coordinates $z = x+iy$, $w = u + i v$, we can reduce to 3-dimensions by making all the fields independent of $y$.  Once we do so, if we identify $\d x$ with $\d \zbar$, we find a theory whose fields in the BV formalism are given by 
\begin{align}
\alpha_{3d} &\in \cinfty(\R \times \C)[\d y, \d \overline{w}] \otimes V \\
\beta_{3d} & \in \cinfty(\R \times \C)[\d y, \d \overline{w}] \otimes V^\ast \d w[1]
\end{align}
with action functional
\begin{equation}
\int \beta_{3d} \d \alpha_{3d} = \int \beta_{3d} (\d y\partial_y + \d \overline{w} \partial_{\overline{w}}) \alpha_{3d}. 
\end{equation}
This is precisely the action we wrote down before. We thus conclude that theory we introduced earlier is a supersymmetric sector of a $3d$ ${\cal N}=2$ scalar field theory.  We have already argued directly that a twist of $3d$ ${\cal N}=2$ pure gauge theory (with a Chern-Simons term) is given by topological Chern-Simons theory itself. It follows that the Chern-Simons matter theory which is the main object of study in this paper is a twist of a $3d$ ${\cal N}=2$ gauge theory with matter and with a Chern-Simons term.   

This analysis also shows that introducing a superpotential term in the matter sector of a  $3d$ ${\cal N}=2$ gauge theory will have the effect of introducing a superpotential term (the dimensional reduction of the one described above) in the Chern-Simons matter theory that describes the twist. (In comparing to the A-model from section 4, one should recall that the complex structure associated to THF used in this section, is different from the complex structure of the $A$-model on $X = X_1 \times X_2$. They differ by a hyper-Kahler rotation of $X_2$, which relates any complex moduli in ${W}$ in \eqref{supp} to Kahler moduli of the $A$-model on $X$, in its natural complex structure. See section \ref{s:rot} for more discussion.)
\section{Examples:   $S^2\times S^1$ and $S^3$}
In this section we consider two sets of examples corresponding to the 3-manifolds $S^2\times S^1$ and $S^3$ with a choice of THF.
For both of these examples we compute the partition function and show that it is equivalent to the partition function of ${\cal N}=2$ supersymmetric theories on the corresponding manifold. For the example of $S^2 \times S^1$ we use operator formulation to rigorously compute the partition function, reaffirming the answer we had anticipated based on localization. For $S^3$ we compute the partition function from two different perspectives, one coming from topological string theory, and the other, from ${\cal N}=2$ gauge theory. The agreement between them is highly non-trivial (even if we take the mass of the bifundamental to infinity); it becomes manifest only at the level of the final answer.

\subsection{$S^2\times S^1$ Example}
In this section, we use a state-operator correspondence to rigorously derived the claimed partition function of our theory on $S^2 \times S^1$ with a particular THF.

\begin{figure}[!ht]
\centering
\includegraphics[height = 3 in]{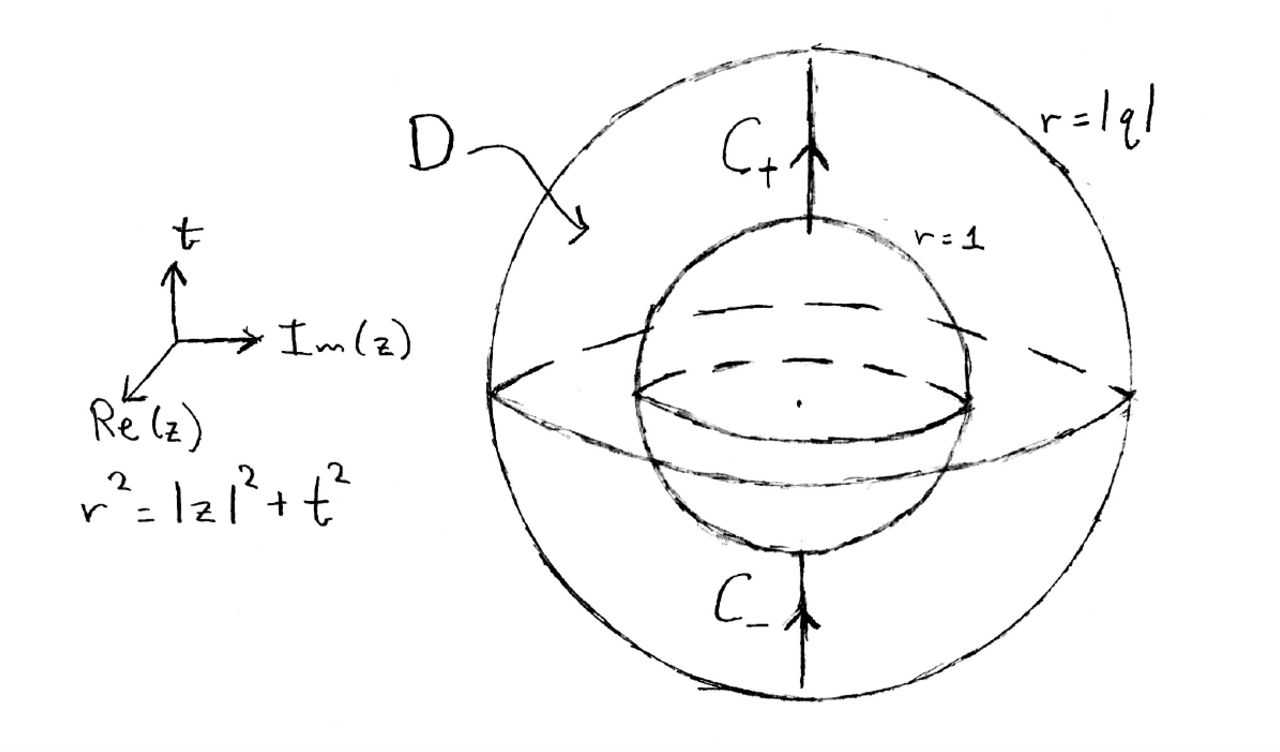}
\caption{The fundamental domain $D$ is the region between the spheres of radius $1$ and radius $\abs{q}$ in $\C \times \R$. The foliation is by the vertical vector field $\del_t$. The inner sphere is glued to the outer sphere by scaling up by $\abs{q}$ and rotating in the $z$-plane by $\arg(q)$. We have also indicated the two closed leaves $\mathcal{C}_\pm$.}
\label{fig:S2xS1}
\end{figure}

To construct this THF, we take the standard coordinates $(z, t)$ on $\mathbb{C} \times \mathbb{R} \smallsetminus (0, 0)$, and identify points by
\[ (z, t) \sim (q z, \,\lvert q \rvert t), \]
where $q \in \mathbb{C}$ with $\lvert q \rvert \neq 1$. We write $M_q$ for the quotient space. For simplicity we assume $\lvert q \rvert > 1$. In this case, a fundamental domain is given by
\[ D = \left\{(z, t) \in \mathbb{C} \times \mathbb{R},\ 1 \leq \sqrt{ \lvert z \rvert^2 + t^2} \leq \lvert q \rvert \right\}. \] Note that $M_q$ is isomorphic to $S^2 \times S^1$ as a manifold, where $S^2$ is the unit sphere in $\mathbb{C} \times \mathbb{R}$ and $S^1$ is the radial circle connecting the inner and outer boundaries of $D$. See Figure \ref{fig:S2xS1} for an image of the fundamental domain $D$.

\subsubsection{Partition function from closed orbits}

We would like to compute the partition function
\[ Z_{matter}(M_q), \]
on the compact manifold $M_q$ in the presence of a flat Chern-Simons background field $A$. The argument in Section \ref{integrate_out_matter} suggests that we should sum contributions from each closed orbit of the THF on $M_q$, of which there are two, given in the domain $D$ by
\[\mathcal{C}_+ = \left\{(0, t) \in D,\ t > 0 \right\}, \quad \mathcal{C}_- = \left\{(0, t) \in D,\ t < 0 \right\}. \]
Both of these are oriented in the direction of increasing $t$, and so $\mathcal{C}_+$ runs from the inner boundary of $D$ to the outer, while $\mathcal{C}_-$ runs from the outer boundary to the inner. As we go around $\mathcal{C}_\pm$, we have that $z \to q^{\pm 1} z$, and further, if the holonomy of $A$ around $\mathcal{C}_+$ is $U$, then the holonomy of $A$ around $\mathcal{C}_-$ is $U^{-1}$, since $A$ is flat and $\mathcal{C}_+$ may be deformed to $\mathcal{C}_-$ with the orientation reversed. Thus, altogether, we expect the partition function to be
\begin{equation}\label{part_fun_Mq} Z_{matter}(M_q) = \prod_{n \geq 0} (1 - q^n U) (1 - q^{-n} U^{-1}),\end{equation}
using (\ref{oo}).

\subsubsection{State-operator correspondence}

To prove that (\ref{part_fun_Mq}) is correct, we will work in the Hilbert space formalism. We associate a Hilbert space to any two-manifold $Y$ together with a germ of a THF on $Y \times (-\epsilon, \epsilon)$. In particular, let $\mathcal{H}$ be the Hilbert space of states on the unit sphere $S^2 \subset \mathbb{C} \times \mathbb{R}$ with the induced germ of a THF from the embedding. This is the Hilbert space arising from radial quantization. From the fundamental domain $D$, we see that
\[ Z_{matter}(M_q) = \text{Tr}_\mathcal{H} \left( \mu_q^{-1} \mu_U^{-1} e^{- \beta H} \right) = \text{Tr}_\mathcal{H} \left( \mu_q^{-1} \mu_U^{-1} \right), \]
since $H = 0$, where $\mu_U$ is the action of the holonomy $U$ of our flat connection around $S^1$ and where $\mu_q$ is the action on $\mathcal{H}$ induced by $(z, t) \to (q z, \lvert q \rvert t)$. We take inverses because we are evolving along $D$ from the inner boundary to the outer, and then gluing back to the inner boundary by $(z, t) \to (q^{-1} z, \lvert q \rvert^{-1} t)$ as well as a gauge transformation given by the inverse of $U$.

To compute this trace, we use a state-operator correspondence. Note that the punctured unit ball
\[ B = \left\{ (z, t) \in \mathbb{C} \times \mathbb{R},\ 0 < \sqrt{ \lvert z \rvert^2 + t^2} \leq 1 \right\}, \]
is isomorphic as a manifold with THF to the half-infinite cylinder,
\[ B \cong S^2 \times (-\infty, 0], \]
and thus, since our theory is topological, we have that states in $\mathcal{H}$ are the same as local operators inserted at the origin in $\mathbb{C} \times \mathbb{R}$. Further, we have that
\[ \mu_q \lvert \mathcal{O} \rangle = q^n \lvert \mathcal{O} \rangle, \quad \Leftrightarrow \quad \mathcal{O}(0) \to q^n \mathcal{O}(0) \text{ when } (z, t) \to (q z, \lvert q \rvert t). \]

\subsubsection{Local operators}

To compute the space of local operators, we use the BV-BRST formalism discussed in Section \ref{BV-BRST}. Recall that we have fields $\Phi, \Psi$ as defined in (\ref{BV fields}), which on $\mathbb{C} \times \mathbb{R}$ are given by
\begin{equation*}
\begin{split}
\Phi &\in \Omega^{1, *}(\mathbb{C}) \otimes \Omega^*(\mathbb{R}) \otimes R, \\
\Psi &\in \Omega^{0, *}(\mathbb{C}) \otimes \Omega^*(\mathbb{R}) \otimes R^*.
\end{split}
\end{equation*}
Further, since we are assuming a flat background connection in this section, we may set $A = 0$ on $\mathbb{C} \times \mathbb{R}$ by a gauge transformation, so the BRST differential is just the de-Rham differential (or the de-Rham differential followed by a projection).

Thus, we may immediately read off the BRST cohomology of local operators. The only fields which survive in cohomology are $\phi$ and $\chi$, which must satisfy the equations
\[ \partial_{\overline{z}} \phi_z = \partial_t \phi_z = 0, \quad \partial_{\overline{z}} \chi = \partial_t \chi = 0. \] Thus, all local operators are obtained by taking products of $\partial_z^n \phi_z(0)$ and $\partial_z^n \chi(0)$. Finally, under the action $(z, t) \to (q z, \lvert q \rvert t)$, the operators transform as
\[ \partial_z^n\phi_z(0) \to q^{- n - 1} \partial_z^n\phi_z(0), \quad \partial_z^n\chi(0) \to q^{-n} \partial_z^n\chi(0). \]

\subsubsection{Partition function from local operators}

With the results of the previous section, we may compute the partition function on $M_q$. We compute, using the fermionic nature of the ghost field $\chi$,
\begin{align*}
Z_{matter}(M_q) &= \text{Tr}_\mathcal{H} \left( \mu_q^{-1} \mu_U^{-1} \right) \\
&= \prod_{n \geq 0} \left( \sum_{k \geq 0} \langle (\partial_z^n \phi_z)^k \rvert \mu_p^{-1} \mu_U^{-1} \lvert (\partial_z^n \phi_z)^k \rangle \right) \\
&\quad \times \left( \sum_{j = 0, 1} \langle (\partial_z^n \chi)^j \rvert \mu_p^{-1} \mu_U^{-1} \lvert (\partial_z^n \chi)^j \rangle \right) \\
&= \prod_{n \geq 0} \left(1 + q^{n + 1} U^{-1} + q^{2(n + 1)} U^{-2}+ \cdots \right) \left(1 - q^n U \right) \\
&= \prod_{n \geq 0} \frac{1 - q^n U}{1 - q^{n + 1} U^{-1}} \\
&= \prod_{n \geq 0} (1 - q^n U)(1 - q^{-n} U^{-1}),
\end{align*}
as desired. In the last line, we used the (formal) identity
\begin{align*}
\log \prod_{n \geq 0} \frac{1}{1 - q^{n + 1} x} &= 
 \sum_{n \geq 0} \sum_{m \geq 1} \frac{q^{m (n + 1)} x^m}{m} \\
&= \sum_{m \geq 1} \frac{x^m q^m}{m (1 - q^m)} = - \sum_{m \geq 1} \frac{x^m}{m (1 - q^{-m})} \\
&= - \sum_{m \geq 1} \sum_{n \geq 0} \frac{x^m q^{- m n}}{m} 
\\
&= \log \prod_{n \geq 0} (1 - q^{-n} x).
\end{align*}

\subsection{$S^3$ Example}

Our second example is $M=S^3$. In this case, $M$ admits a single complex family of THF's, parameterized by one complex number $b\in {\mathbb C}^{\times}$.  We will first explain how this structure arizes in topological string on $X=T^*S^3$. Then, we will explain how to compute the partition function of $N$ 3-branes on the $S^3$ with one 5-brane, supported on a coisotropic manifold $Y$, which depends on $b$.  The 3-5 brane setup conjecturally realizes the  $U(N)$ level $k$ Chern-Simons theory, with single fundamental matter multiplet, as in section two, with the choice of THF corresponding to $b$. Finally, we will explain how to compute the partition function of this system, and show it agrees with that of the corresponding 3d ${\cal N}=2$ theory, assuming the same THF.

\subsubsection{A-model on $T^*S^3$}

The cotangent bundle to $S^3$, 
$$X=T^*S^3$$ 
is a hypersurface

\beq\label{X}
X: \qquad  xy -zw =\mu,
\eeq
in ${\mathbb C}^4$ with coordinates $x,y,z,w$. We will assume $\mu \in {\mathbb R}_{>0}$ to be real and positive. 
The real symplectic form of $X=T^*S^3$ is
$$\omega = {i}(dx \wedge d{\bar x}+dy \wedge d{\bar y} +dz \wedge {d \bar z} + dw \wedge d {\bar w}),
$$
and the holomorphic $(3,0)$ form is
$$
\Omega = dx \wedge dy \wedge {dz / z}
$$
\subsubsection{Coisotropic Brane}
A coisotropic brane $Y$ in this geometry is given by the vanishing set 

$${\cal H}(x,y,z,w)=m
$$ 
of the function 

$$
{\cal H}(x,y,z,w)= {1\over b} (|x|^2-|y|^2) + b(|z|^2-|w|^2) ,
$$
 where we take $m$ and $b$  to be real and positive. It is easy to work out that $v$ is given by

\beq\label{vector}
v = {i \over b} \left(x {\partial \over \partial x} - y {\partial  \over \partial  y}\right)+ i b \left(z {\partial  \over \partial { z}}  - w {\partial \over \partial  w} \right) + c.c.
\eeq
This rotates $x$ and $y$, and $z$ and $w$ by opposite phases, by amounts proportional to $b$ and $1/b$, in such a way that one stays in the level set of the Hamiltonian ${\cal H}.$
The corresponding flux is given from \eqref{Fe} by
$$
F={\rm Im}(i_v \; \Omega)
$$
or
\beq\label{fluxc}
F= i (\,b\, dx \wedge dy - {1\over b}\,dz \wedge d w) - c.c.
\eeq
For any  positive $b$, $Y$ is a coisotropic submanifold of $X$, of  topology $S^3 \times {\mathbb R}^2$.

\subsubsection{3-5 Brane System}

Now, consider adding $N$ Lagrangian branes on the $S^3.$ The $S^3$ in the base is the hypersurface in $X$ given by

\beq\label{S3}
x=\bar{y}, \qquad z=-{\bar w}.
\eeq 
$S^3$ is Lagrangian as the symplectic form vanishes $\omega|_{S^3} =0$ on it.
The restriction of the coisotropic brane flux $F$ in \eqref{fluxc}  to the $S^3$ vanishes as well,
$$
F|_{S^3} =0.
$$
As we showed in section \ref{s:TFHt} this implies that the coisotropic brane introduces a transverse holomorphic foliation on the $S^3$.  The THF structure on the Lagrangian branes on the $S^3$ is induced through the bi-fundamental matter sector. 

Namely, if we write the $S^3$ as 

\beq\label{S3a}
|x|^2 + |z|^2 =\mu,
\eeq
coming from restricting \eqref{S3} to $X$, then the induced THF has "time" direction generated by the vector field $v$ in \eqref{vector}. Restricted to the $S^3$,  \eqref{vector} reads

\beq\label{vectorc}
v = {i \over b} \left(x {\partial \over \partial x} - {\overline x} {\partial  \over \partial  {\overline x}}\right)+ i b \left(z {\partial  \over \partial { z}}  - {\overline z} {\partial \over \partial  {\overline z}} \right),
\eeq
up to a factor of $2$. For finite time $t$, this takes a point on the $S^3$ with coordinates $(x,z)$  to
\beq\label{orbitaction}
(x,z) \;\in\; S^3 \qquad \rightarrow \qquad (x\,e^{i t /b},  z\,e^{ ib t}) \;\in\; S^3.
\eeq

The 3-5 strings, as explained in section 3, lead to a matter fields $(\phi, \eta)$ in $(R, R^c)$ representations of the $U(N)$ gauge group, where $R$ is the fundamental representation of $U(N)$. We could have easily considered $M$ coisotropic fivebranes on $Y$ instead of one, which would give $M$ copies of this, and $U(M)$ flavor symmetry. For the rest of this section, and for simplicitly only, we will restrict to the $M=1$ case.

\subsubsection{Integrating out matter}

Integrating out the bifundamental matter fields $(\phi, \eta)$ corresponds to summing over one loop open string diagrams.
Only the diagrams associated to closed orbits of THF make a non-zero contribution, as we found in section 2. This is natural from the A-model perspective. 

In the A-model, we consider holomorphic maps with boundaries on the branes. 
Boundaries of holomorphic maps ending 
on the coisotropic brane have to fall on the orbits of the vector field $v$, because only in that case is the surface spanned by the normal vector field to the brane and the boundary holomorphic. To get finite action contributions, we need to further restrict to the closed orbits of $v$.

For $b$ rational, closed orbits come in families. For example, for $b=1$, the closed orbits correspond to fibers of the Hopf fibration of the $S^3$, so there is an $S^2$ worth of them. For irrational $b$, however, there are only two closed orbits on the $S^3$. 
The closed orbits are the two circles at 

$$x=0 \;\;\textup{     and       }  \;\; z=0,
$$
each of which is fixed by \eqref{orbitaction}.

Pick now local coordinates on the $S^3$ adapted to the neighborhood of one of these two closed orbits. Near the orbit at $z=0$, say,  the  Lagrangian of bifundamental fields takes the form
$$
\int  d\theta_x \,d^2z \;\eta_{\overline {z}} \Bigl(  {1\over b }( \partial_{\theta_x}-i A_{\theta_x}) + m+ {i b} {z}{\partial\over \partial z}\Bigr)\phi_{ z},
$$ 
where $\theta_x$ is the phase of $x$ and parameterizes the closed leaf.
Thus, in effect, contribution of string states along the $z=0$ orbit on the $S^3$ is a tower of particles, coming from modes of supported on the $S^1$, with masses
$$
mb + ikb^2, \qquad k\in {\mathbb Z}_{\geq 0}
$$
transforming in the bifundamental representation of the gauge group. The operator induced from integrating out the bifundamental strings localized at $x=0$ is 
$k$,
\beq\label{z}
Z_{matter}({\cal C}_{z=0}; U_{z=0}) =\prod_{k=0}^{\infty} \det(1 - e^{-2\pi mb-i kb^2} U_{z=0}).
\eeq
which agrees with that in section $2$ if we put 
$$q_{z=0} = e^{i b^2}.$$
Above, $U_{z=0}$ is the holonomy of the Chern-Simons gauge field along the circle at $ z=0$ in $X$.
From the other closed orbit, at $x=0$,by analogous consideration  we get 
\beq\label{x}
Z_{matter}({\cal C}_{x=0}; U_{x=0}) =\prod_{k=0}^{\infty} \det(1 - e^{-2\pi m/b-i k/b^2} U_{x=0} ).
\eeq
corresponding to
$$q_{x=0} = e^{i/ b^2}.$$

Our derivation strongly suggests that the topological string on $T^*S^3$  in this setting is described by Chern-Simons/matter system from section 2. 

\subsubsection{A-model partition function}

%
We have argued that topological string theory on $T^*S^3$ with $N$ branes on the $S^3$ and a single coisotropic brane on $Y$ leads to of $U(N)$ Chern-Simons theory on $S^3$ with the $(\eta, \phi)$ system on $S^3$, and THF induced by the coisotropic brane.

The A-model partition function on $T^*S^3$ with branes is computed by $U(N)$, level $k$ Chern-Simons theory on $S^3$, with insertion of operators on the $S^3$, 

\beq\label{S3p}
Z_{top}= \langle Z_{matter}({\cal C}_{z=0}; U_{z=0}) \cdot Z_{matter}({\cal C}_{x=0}; U_{x=0})\rangle_{S^3},
\eeq
coming from integrating out matter, in \eqref{z} and in \eqref{x}. This is a particular realization of  eqn. \eqref{eoo} from section 2. 

The line operators in \eqref{z} and \eqref{x} lead to Wilson-lines in Chern-Simons theory along the two unknots in the $S^3$ which are Hopf-linked. To see that, one views the $S^3$ in \eqref{S3a} as a $T^2$ fibration by phases of $x$ and $z$ over an interval. The two finite orbits at $x=0$ and $z=0$ correspond to the $(1,0)$ and the $(0,1)$ cycles of the $T^2$. They are linked, and the corresponding link is the Hopf link (see for example \cite{MM}).

The operators \eqref{z} and \eqref{x} are gauge invariant, and thus have an expansion in terms of traces of the $U(N)$ gauge group in various representations, which one can obtain using:

$$
\prod_k \det(1-U V_k) = \sum_R (-1)^{|R|} {\rm Tr}_R(U) {\rm Tr}_{R^T}(V) 
$$
where 
$$V = {\rm diag}(V_1, \ldots, V_k, \ldots)$$ 
is a matrix with eigenvalues $V_k$. The sum goes over all Young diagrams, this translates to a sum over all $SU(N)$ representations. For the fixed point at $x=0$, 

$$V_k (m,b)= m/b + ik/b^2, \qquad k\in {\mathbb Z}_{\geq 0}
$$
Thus, the guts of the computation of the partition function in \eqref{S3p} involves computing
the Chern-Simons path integral with insertion of
\beq\label{Hopf}
 {\rm Tr}_R U \cdot  {\rm Tr}_Q U
\eeq
corresponding to two Wilson loops linked into a Hopf link, and colored by arbitrary representations
$R$ and $Q$. As explained in \cite{Witten:1988hf}, this is computed by 
$$
S_{RQ} =\langle  {\rm Tr}_R U \cdot  {\rm Tr}_Q U  \rangle 
$$ 
where $S$ is the $S$-matrix of the corresponding $SU(N)_k$ WZW model, and we take its $Q$, $R$ matrix element.

\subsubsection{A subtlety - framing in CS theory}

There is an important technical subtlety that enters any computation of knot invariants in Chern-Simons theory. To fully specify any knot observables in Chern-Simons theory, we need to specify the framing of the knot \cite{Witten:1988hf}. This affects our computation as well.  Framing is a choice of a normal vector field at every point on the knot $K$. In effect this pairs a knot with a copy $K'$ of itself translated by the normal vector, and framing can be taken to me the linking number of $K$ and $K'$. For knots in $S^3$ there is a  canonical framing, corresponding to asking the liking number of the knots $K$ and $K'$ to be zero. 

In our problem, the framing vector field is naturally provided by $v$. Viewed as a vector on the tangent space to the $S^3$, $v$ is given by

$$
v|_{S^3} = {i   x/b} {\partial \over \partial x} + i b \, z {\partial  \over \partial { z}} + cc 
$$
The knots we consider are the two orbits of $v$ at $x=0$ and $z=0$. Considering $v$ in the neighborhood of the two knots naturally defines a  vector `tangent to knot $K'$', at least infinitesimally.
For a knot $K$ along the $x=0$ circle in the $S^3$, the tangent vector to the knot is $ i z {\partial  \over \partial { z}} +cc  = {\partial  \over \partial \theta_{ z}} $ is tangent to the knot, and we shift it by $ i/b^2 x {\partial  \over \partial { x}}+cc = {1/b^2} {\partial  \over \partial \theta_{ x}}$. This corresponds to $1/b^2$ units of framing, at least formally since $b$ is allowed to be irrational.
Similarly, for the knot at $z=0$ we get $b^2$ units of framing, so all together, the contribution to the partition function of Wilson loops colored by $R$ and $Q$ is modified to\footnote{
Changing a framing by $n$ and $m$ units for the two knots, we would
get
$$
T_R^mS_{RQ}T_Q^n
$$ 
instead, where $T_{RQ} = T_R \delta_{RQ}$ is the $T$ matrix of the WZW model.}

$$
T_R^{1/b^2}\, S_{RQ}\,T_Q^{b^2} = \langle  {\rm Tr}_R U \cdot {\rm Tr}_Q U  \rangle_{framed}
$$ 
Combining all the factors, the partition function equals:

\beq\label{top3}
Z_{top} = \sum_{R, Q} \; (-1)^{|R|+|Q|} \;T_R^{1/b^2} S_{RQ} T_R^{b^{2}}  \; {\rm Tr}_{R^T} V(m,1/b) {\rm Tr}_{Q^T} V(m,b)
\eeq
where the sum runs over all Young diagrams, and taking $m\gg 0$ ensures its convergence.

\subsubsection{${\cal N}=2$ gauge theory on $S^3$}

Take now the ${\cal N}=2$ $SU(N)$ gauge theory with a single chiral multiplet in fundamental representation of mass $m$, and 
with Chern-Simons level $k$.  We study the theory on the $S^3$ with THF and parameter $b$. The $S^3$ with such a structure can be modeled \cite{Dumitrescu:2016ltq} by a squashed three sphere 
$$b^2(|x_1|^2+|x_2|^2)+{b^{-2}}(|x_3|^2+|x_4|^2)=1,$$
with squashing parameter $b$, and $x_i\in {\mathbb R}$. The partition function of the ${\cal N}=2$ theory
was studied by Hama, Hosomichi and Lee in \cite{Hama:2010av}.

We will now show that the partition function $Z_{{\cal N}=2}$ they found equals that of the topological string system on $T^*S^3$, given in \eqref{top3}.  The Chern-Simons level $k$ of the ${\cal N}=2$ theory is equal to the level of Chern-Simons in topological string, and the real masses of chiral multiplets agree as well. The levels are related to the topological string coupling $g_s$ in the usual way \cite{Witten:1992fb}, by $g_s=2\pi i /(k+N)$. 

\subsubsection{${\cal N}=2$ partition function}

The partition function presented in \cite{Hama:2010av} is a matrix integral, over the eigenvalues of the scalar $x$ in the vector multiplet. The contribution of the ${\cal N}=2$ vector multiplet is given in eqn 5.33 of their paper (the denominator of that equation cancels with a another factor, as they explain)

$$
\prod_{\alpha \in \Delta_+}{\sinh(\pi b \alpha \cdot x)\sinh(\pi b^{-1} \alpha \cdot x)}
$$
where $\Delta_+$ is the set of all positive roots. The contribution of the chiral multiplet is given by the double sign function:
$$
\prod_{i=1}^N s_b\left(i\frac{Q}{2}+m-x_i\right) 
$$
where $Q=b+b^{-1}$. The double sine function is defined by
$$s_b(x)= \prod_{\ell,n\geq 0}{\frac{\ell b+n b^{-1} +{Q\over 2} -i x}{\ell b+n b^{-1} +{Q\over 2} +i x}}
$$
This can be rewritten as (up to a constant prefactor)
%
$$
s_b(x) =  e^{- i \frac{\pi}{2}x^2} \prod_{n\geq 0}(1+p^{n-{1\over 2}}\; e^{2\pi b x})(1+{\hat p}^{n-{1\over 2}}\; e^{2\pi b^{-1}x })
$$
where 
\beq\label{ps}
p = e^{2 \pi i b^2}, \qquad {\hat p}=e^{2 \pi i  /b^{2}}.
\eeq
The full partition function of the ${\cal N}=2$ theory, taking into account the Chern-Simons level $k$ is \cite{Hama:2010av}

\begin{align}\label{inf}
{1\over N!} \int d^N x \prod_{\alpha \in \Delta_+}&{\sinh(\pi b \alpha \cdot x)\sinh(\pi b^{-1} \alpha \cdot x)}
e^{ \pi i k' \sum_i  x_i^2} \cr \times
\prod_{i=1}^N  &\prod_{n\geq 0}(1-p^{n}\; e^{2\pi b (m'-x_i)})(1-{\hat p}^{n}\; e^{2\pi b^{-1}(m'-x_i) })
\end{align}
where $k'=k-1/2$ and $m'$ is $m$ shifted by terms (depending on $k$ and $b$) needed to bring the  quadratic term in the gaussian to the canonical form we used above.

\subsubsection{Evaluating the integral}
To compute the integral, we proceed as follows. (The computation is necessarily technical, involving some of the techniques developed in \cite{marino, MM}. The reader not interested in the details may want to skip to the final answer.) Firstly, using
$$
\prod_{ij}(1-x_i y_j) = \sum_{R}  (-1)^{|R|} {\rm Tr}_{R^T} x\; {\rm Tr}_{R} y
$$
where  ${\rm Tr}_{R} y$ is the trace in representation $R$ of a matrix $y$ with eigenvalues $y_i$,
we can rewrite
$$
\prod_{i=1}^N  \prod_{n\geq 0}(1-p^{n}\; e^{2\pi b (m'-x_i)}) =\sum_R (-1)^{|R|} {\rm Tr}_{R^T} V(m, b) {\rm Tr}_{R}(e^{2\pi b x})
$$
where $V(m,b)$ is a diagonal matrix with eigenvalues
$$
V(m, b) = \exp(2\pi b m') \cdot  {\rm diag}(1, p , p^2, \ldots)
$$
where $p = e^{2 \pi i b^2}$ from \eqref{ps}. Similar result holds for the other factor, with $b$ and $b^{-1}$ exchanged.
The relevant part of the integral to compute is thus
$$
{1\over N!} \int d^N x \prod_{\alpha \in \Delta_+}{\sinh(\pi b \alpha \cdot x)\sinh(\pi b^{-1} \alpha \cdot x)}
e^{ \pi i k' \sum_i  x_i^2} \;\; {\rm Tr}_R (e^{2\pi b x}){\rm Tr}_Q (e^{2\pi b^{-1} x})
$$

Second, using
$$
\prod_{\alpha \in \Delta_+}\sinh(\pi b \alpha \cdot x)=\sum_{\omega \in S_N} (-1)^{\omega} e^{2 \pi b \omega(\rho_N)\cdot x }
$$
and
$$
\prod_{\alpha \in \Delta_+}\sinh(\pi b \alpha \cdot x){\rm Tr}_R  (e^{2\pi b x})=\sum_{\omega \in S_N} (-1)^{\omega} e^{2 \pi b \omega(\rho_N+R)\cdot x }
$$
we can reduce the integral to a gaussian:
\begin{align}
{1\over N!} \int d^Nx \sum_{\omega, \hat \omega \in S_N} &(-1)^{\omega+\hat \omega} e^{2 \pi b \omega(\rho_N+R)\cdot x }
e^{2 \pi b^{-1} \omega(\rho_N+Q)\cdot x }
e^{ \pi i k' \sum_i  x_i^2} \cr
\;\; 
 \sim {1\over N!} \sum_{\omega, \hat \omega \in S_N} &(-1)^{\omega+\hat \omega} e^{{\pi i\over  k'}( b \omega(\rho_N+R)+b^{-1} \omega(\rho_N+Q))^2}\cr
= q^{b^2 C_2(R)/2} &S_{RQ}(q) q^{b^{-2} C_2(Q)/2}
\end{align}
Above, $S_{RQ}$ is the standard $S$-matrix of $SU(N)$ Chern-Simons theory at the effective level $k'$,
$$
S_{RQ}={1\over N!} \sum_{\omega \in S_N} (-1)^{\omega} q^{  \omega(\rho_N+R)\cdot (\rho_N+Q))}
$$
where
$$
q = e^{2 \pi i \over k'}
$$
and $ q^{C_2(R)/2} = T_R$ is the standard framing vector. Note that it gets raised to the power either $b^2$ or $b^{-2}$.

All together, by computing the integral in \eqref{inf}, we obtained a concrete, numerical expression
\beq\label{N2p}
Z_{{\cal N}=2}= \sum_{R, Q} (-1)^{|R|+|Q|} \; T_R^{b^2} S_{RQ}(q) T_R^{b^{-2}} \;  {\rm Tr}_{R^T} V(b, m) {\rm Tr}_{Q^T} V(1/b, m)
\eeq
for the ${\cal N}=2$ partition function.

\subsubsection{Summary }

At this point, we can simply compare by inspection, the topological string partition function $Z_{top}$ from \eqref{top3}, with the ${\cal N}=2$ partition function as written in  \eqref{N2p}. They are manifestly equal,

$$
Z_{{\cal N}=2} = Z_{top},
$$
with the gauge group ranks, the THF parameter $b$, the level $k$ and the mass $m$ trivially identified. This is as expected, since as we argued in sections 3-5, both are described by the same, partially topological Chern-Simons matter theory, given in section 2.
\section{Possible Condensed Matter Application}
By now it is a well known fact that topological field theories can arise as the IR description of condensed matter systems.  Chern-Simons theory in particular has been proposed as an effective description of quantum Hall effect.  In this context, the anyons are viewed as external sources that can be coupled to the Chern-Simons theory.  The anyons would transform in some representation of the gauge group and appear as infinitely massive sources which lead to insertion of Wilson loop observables in the CS theory.  

In a physical system, anyons are dynamical and have finite mass. However, coupling the matter in the usual way to CS theory would completely ruin its topological nature.  In the setup we have been discussing, we have managed to preserve the topological nature of CS theory while coupling to matter at the expense of requiring the underlying 3-manfiolds to admit a THF structure.  The question is whether such a theory can arise as an IR limit of a condensed matter system? 
Indeed it has been proposed in \cite{Kong:2014qka, Freed:2016rqq} (for a review see \cite{WenTalk}) that there could exist topological theories that are not strictly topological, but that they admit a splitting of temporal and spatial directions.  These are called topological theories of class H.  

Our theory, which requires a THF structure, realizes this idea, because locally we would need a splitting of the space to $(z,t)$ where we can identify $t$ as the temporal direction.  More precisely, even though the vector field $\partial /\partial t$
may not be globally well defined, at each point there is a well defined direction which can be identified with the diffeomorphism generated by $\partial /\partial t$. 
This direction can be identified as the kernel of $dz$ and $d{\overline z}$ at each point.  So in this sense our topological theory can be viewed as a topological theory of class $H$. 

In the context of CS theory we can view our matter fields as anyonic fields of mass $m$.  To be concrete, suppose we consider the geometry
$\mathbb{R}^2\times S^1$ where we view $\mathbb{R}^2$ as the space and $S^1$ as putting the theory in a thermal bath where the radius of $S^1=\beta=1/kT$ and as we go around $S^1$ we rotate the ${\mathbb R}^2$ by $\theta$.  So this partition function can be viewed as
$${\rm Tr}\  {\rm exp}[-\beta H +i\theta J]$$
We can view the $\theta$ term as putting a chemical potential for rotation $J$ on the plane:  $\beta \mu =-i\theta$, i.e.
$${\rm Tr}\  {\rm exp}[-\beta (H +\mu J)]$$
As discussed in this paper in the context of our model this is easily computed and yields
$$Z= \langle \prod_n (1-q^n  e^{-\beta m} U)\rangle .$$
where $q={\rm exp}(-\beta \mu)$ and the integrating out of the anyonic field yields the above insertion in the Chern-Simons theory, where $U$ is the holonomy of the gauge field along the $S^1$.  It would be interesting to see if such a term can be connected to theoretical and/or experimental setup of QHE.

\section*{Acknowledgements}

We would like to thank Clay Cordova, Thomas Dumitrescu and Anton Kapustin for very useful discussions. We are especially grateful to Simeon Hellerman and
Daniel Jafferis who participated in early stages of this work.

M.A.\ is supported by NSF grant 1521446, by the Berkeley Center for
Theoretical Physics and by the 
Simons Foundation as a Simons Investigator. K.C. is supported by the NSERC Discovery Grant program and by the Perimeter Institute for Theoretical Physics. Research at Perimeter Institute is supported by the Government of Canada through Industry Canada and by the Province of Ontario through the Ministry of Research and Innovation. J.M. and C.V. gratefully acknowledge support from the Simons Center for Geometry and Physics, Stony Brook University, where some of the research for this paper was performed during the 2016 Simons Summer Workshop. C.V. is supported in part by NSF grant PHY-1067976.

\end{document}